\RequirePackage[l2tabu, orthodox]{nag}
\RequirePackage{snapshot}

\documentclass[9pt,onecolumn]{extarticle}

\makeatletter
\if@twocolumn
  \usepackage[dvips,letterpaper,top=0.5in, bottom=0.5in, left=0.75in, right=0.5in,includefoot,heightrounded]{geometry}
\else
  \usepackage[dvips,letterpaper,margin=1in,includefoot,heightrounded]{geometry}
\fi

\usepackage{srcltx}

\usepackage{abstract}
\usepackage[russian,portuges,english]{babel}

\iflanguage{portuges}
    {\newcommand{\keywordname}{Palavras-chaves}}
    {\newcommand{\keywordname}{Keywords}}

\usepackage{lineno}

\usepackage{amsmath}
\usepackage{amssymb,amsfonts}

\usepackage{psfrag}
\usepackage{enumerate}
\usepackage{subfigure}
\usepackage{epsfig}
\usepackage{color}
\usepackage[usenames,dvipsnames]{xcolor}

\usepackage{booktabs}

\usepackage{fullpage}
\usepackage{setspace}
\usepackage{flushend}
\usepackage{multicol}
\usepackage{multirow}
\usepackage{url}
\usepackage[normalem]{ulem}
\usepackage{bm}

\usepackage{booktabs}

\usepackage{setspace}
\usepackage{flushend}

\usepackage{hyperref}

\usepackage{MnSymbol}

\makeatletter

\newcommand{\printtitle}{%
\makeatletter
\if@twocolumn

\twocolumn[%
  \maketitle
  \begin{onecolabstract}
    \myabstract
  \end{onecolabstract}
  \begin{center}
    \small
    \textbf{\keywordname}
    \\\medskip
    \mykeywords
  \end{center}
  \bigskip
]
\saythanks
\else
  \maketitle
  \begin{onecolabstract}
    \myabstract
  \end{onecolabstract}
  \begin{center}
    \small
    \textbf{\keywordname}
    \\\medskip
    \mykeywords
  \end{center}
  \bigskip
  \onehalfspacing
\fi
\makeatother
}

\author{%
C.~J.~Tablada%
\thanks{Signal Processing Group,
 Departamento de Estat\'{\i}stica,
 Universidade Federal de Pernambuco, Recife, PE, Brazil}
\and
T.~L.~T.~da~Silveira%
\thanks{Programa de P\'os-Gradua\c{c}\~ao em Computa\c{c}\~ao,
 Universidade Federal do Rio Grande do Sul, Porto Alegre, RS, Brazil}
\and
R.~J.~Cintra%
\thanks{%
Signal Processing Group,
 Departamento de Estat\'{\i}stica,
 Universidade Federal de Pernambuco, Recife, PE, Brazil
E-mail: \url{rjdsc@de.ufpe.br}}
\and
F.~M.~Bayer%
\thanks{Departamento de Estat\'istica and LACESM,
 Universidade Federal de Santa Maria, Santa Maria, RS, Brazil}
}

\title{%
DCT Approximations Based on Chen's Factorization}

\newcommand{\myabstract}{%
In this paper,
two 8-point
multiplication-free
DCT approximations
based on
the Chen's factorization
are proposed
and their
fast algorithms
are also
derived.
Both
transformations are
assessed in terms of
computational cost,
error energy,
and
coding gain.
Experiments with a JPEG-like image compression
scheme are performed and results are compared with
competing methods.
The proposed low-complexity transforms
are scaled according to Jridi-Alfalou-Meher
algorithm
to effect 16- and 32-point approximations.
The new sets of transformations
are
embedded into an HEVC reference software
to provide a fully HEVC-compliant
video coding scheme.
We show that
approximate transforms
can outperform
traditional transforms
and state-of-the-art methods
at a very low complexity cost.
}

\newcommand{\mykeywords}{%
Approximate DCT,
Chen's factorization,
Fast algorithms,
Image and video compression,
Low-complexity transforms

}

\date{}

\begin{document}

\printtitle

\section{Introduction}

Discrete transforms
are very useful tools
in digital signal processing
and compressing technologies~\cite{Britanak2010,Horadam2010}.
In this context,
the discrete cosine transform (DCT)
plays a key role~\cite{Ahmed1974}
since it is
a practical approximation for
the Karhunen-Lo\`eve transform (KLT)~\cite{Effros2004}.
The KLT has the property of being
optimal
in terms of energy compaction
when the
intended signals
are modeled after a highly correlated
first-order Markov process~\cite{Britanak2010}.
This is a widely accepted supposition for natural images~\cite{Gonzalez2006}.

Particularly,
the DCT type~II (DCT-II) of order~8
is a widely employed method~\cite{Britanak2010},
being adopted in several industry standards
of image and video compression,
such as
JPEG~\cite{Gonzalez2006},
MPEG-1~\cite{Roma2007},
MPEG-2~\cite{mpeg2},
H.261~\cite{h261},
H.263~\cite{h263},
H.264~\cite{Puri2004},
and
the state-of-the-art HEVC~\cite{Pourazad2012}.
Aiming at the efficient computation of
the DCT,
many fast algorithms have been reported
in literature~\cite{Chen1977, Loeffler1989, Suehiro1986, Yuan2006}.
Nevertheless,
these methods usually need
expensive arithmetic operations,
as multiplications,
and an arithmetic on floating point,
demanding major
hardware requirements~\cite{Liang2001}.

An
alternative
to the exact
DCT computation
is the use of DCT approximations
that
employ integer arithmetic only
and
do not require multiplications~\cite{Britanak2010, Horadam2010}.
In this context,
several approximations for the DCT-II
were proposed in literature.
Often
the elements of approximate transform matrices
are defined over
the set~$\mathcal{P}=\{0,\pm\frac{1}{2},\pm 1,\pm 2 \}$~\cite{Haweel2001, Cintra2011}.
Relevant methods
include the
signed DCT (SDCT)~\cite{Haweel2001}
and the Bouguezel-Ahmad-Swamy (BAS)
series of approximations~\cite{Bouguezel2008, Bouguezel2010, Bouguezel2011, Bayer2012}.

Transform matrices with elements in~$\mathcal{P}$
have null multiplicative complexity~\cite{blahut2010fast}.
Thus
their
associate
hardware
implementations
require only
additions and bit-shifting operations~\cite{Bayer2012}.
This fact renders
such
multiplierless approximations
suitable for hardware-software implementations
on devices/sensors
operating at low computational power and
severe energy consumption constraints~\cite{Chang2000, Puri2004, Roma2007}.

In this work,
we aim at the following goals:
\begin{itemize}

\item
the proposition of new
multiplication-free
approximations for the 8-point DCT-II,
based on Chen's algorithm~\cite{Chen1977};

\item
the derivation of fast
algorithms for the introduced
transforms;

\item
a comprehensive assessment
of the new approximations
in terms of
coding and image compression performance
compared to
popular alternatives;

\item
the extension
of the proposed 8-point transforms to
16- and 32-point DCT approximations
by means
of
the scalable recursive method proposed in~\cite{Jridi2015};
and

\item
the embedment of the obtained approximations
into an HEVC-compliant reference software~\cite{HEVCReference}.

\end{itemize}

This paper unfolds as follows.
Section~\ref{s:chen} presents Chen's factorization for the 8-point DCT-II.
In Section~\ref{s:DCT_approximations},
we present two novel
8-point
multiplierless transforms and their fast algorithms.
The proposed approximations are assessed and mathematically
compared with
competing methods
in
Section~\ref{s:assessment}.
Section~\ref{s:image_compression}
provides comprehensive image compression analysis
based on a JPEG-like compression scheme.
Several images are compressed and assessed for quality
according to the approximate transforms.
Section~\ref{s:video_coding}
extends the 8-point transforms
to 16- and 32-point DCT approximations
and considers
a real-world
video encoding scheme
based on these particular DCT approximations.
Final conclusions and remarks
are drawn in Section~\ref{s:conclusion}.

\section{Chen's factorization for the DCT}\label{s:chen}

Chen~\emph{et~al.}~\cite{Chen1977}
proposed
a fast algorithm for the DCT-II
based
on a factorization for the
DCT type IV (DCT-IV).
These two versions of the DCT differ in the sample points
of the cosine function
used in their transformation kernel~\cite{Wang1984,Britanak2010}.
The $(m,n)$-th element of the $N$-point DCT-II and DCT-IV transform matrices,
respectively denoted as
$\mathbf{C}_N^{\mbox{\scriptsize II}}$
and
$\mathbf{C}_N^{\mbox{\scriptsize IV}}$,
are given by:
\begin{align*}
\left[
\mathbf{C}_N^{\mbox{\scriptsize II}}
\right]_{m,n}
=
&
\left[
\sqrt{\frac{2}{N}}
\,\,
c_m
\cos\!\left(\frac{m(2n+1)\pi}{2N}\right)
\right]_{m,n},
\\
\left[
\mathbf{C}_N^{\mbox{\scriptsize IV}}
\right]_{m,n}
=
&
\left[
\sqrt{\frac{2}{N}}
\,
\cos\!\left(\frac{(2m+1)(2n+1)\pi}{4N}\right)
\right]_{m,n}
,
\end{align*}
where $m,n=0,1,\ldots,N-1$ and
\begin{align*}
 c_m=\left\{\begin{array}{cl}
             1/\sqrt{2},& \text{if $m=0$},\\
             1,& \text{if $m\neq 0$}.\\
            \end{array}\right.
\end{align*}

In the following,
let
$\mathbf{0}_N$
the
zero matrix of order $N$,
$\mathbf{I}_N$
the
$N\times N$
identity
and
$\bar{\mathbf{I}}_N$
the counter-identity matrix,
which is given by:
\begin{align*}
\bar{\mathbf{I}}_N=\left[\begin{smallmatrix}
                   \ \ &\ \ &\ \ &\ \ \\
                   0&\cdots&0&1\\[0.05cm]
                   0&\cdots&1&0\\[0.05cm]
                   \vdots&\udots&\vdots&\vdots\\[0.05cm]
                   1&\cdots&0&0\\[0.05cm]
                  \end{smallmatrix}
\right]
.
\end{align*}
In~\cite{Wang1983},
Wang demonstrated that
the 8-point DCT-II matrix
possesses the following factorization:
\begin{equation}
\label{ec:ecuation1}
\mathbf{C}_8^{\mbox{\scriptsize II}}
=
\frac{1}{2}
\,
\mathbf{P}_8
\left[
\begin{array}{cc}
\mathbf{C}_4^{\mbox{\scriptsize II}} & \mathbf{0}_4\\
\mathbf{0}_4&\bar{\mathbf{I}}_4\,\mathbf{C}_4^{\mbox{\scriptsize IV}}\,\bar{\mathbf{I}}_4\\
\end{array}
\right]
\mathbf{B}_8
,
\end{equation}
where
$\mathbf{P}_8$ and $\mathbf{B}_8$ are permutation and pre-addition matrices
given by, respectively:
\begin{align*}
 \mathbf{P}_8=\left[\begin{smallmatrix}
           \ \ &\ \ &\ \ &\ \ &\ \ &\ \ &\ \ &\ \ \\[0.05cm]
           1&0&0&0&0&0&0&0\\[0.05cm]
           0&0&0&0&0&0&0&1\\[0.05cm]
           0&1&0&0&0&0&0&0\\[0.05cm]
           0&0&0&0&0&0&1&0\\[0.05cm]
           0&0&1&0&0&0&0&0\\[0.05cm]
           0&0&0&0&0&1&0&0\\[0.05cm]
           0&0&0&1&0&0&0&0\\[0.05cm]
           0&0&0&0&1&0&0&0\\[0.05cm]
          \end{smallmatrix}
\right],\ \mathbf{B}_8=\left[\begin{array}{rr}
           \mathbf{I}_4&\bar{\mathbf{I}}_4\\
           \bar{\mathbf{I}}_4&-\mathbf{I}_4\\
          \end{array}
\right].
\end{align*}
Additionally,
Chen~\emph{et~al.} suggested in~\cite{Chen1977}
that
the matrix $\mathbf{C}_4^{\mbox{\scriptsize IV}}$
admits the following
factorization:
\begin{equation}
\label{ec:ecuation2}
\mathbf{C}_4^{\mbox{\scriptsize IV}}=\mathbf{Q}\,\mathbf{A}_1\,\mathbf{A}_2 \,\mathbf{A}_3,
\end{equation}
where
\begin{align*}
\mathbf{Q}=\left[\begin{smallmatrix}
           \ \ &\ \ &\ \ &\ \ \\
           1&0&0&0\\[0.05cm]
           0&0&1&0\\[0.05cm]
           0&1&0&0\\[0.05cm]
           0&0&0&1\\[0.05cm]
          \end{smallmatrix}
\right], \;
\mathbf{A}_1=\left[\begin{smallmatrix}
             \ \ &\ \ &\ \ &\ \ \\
             \beta_0&\phantom{-}0&\phantom{-}0&\phantom{-}\beta_3\\[0.05cm]
             0&\phantom{-}\beta_2&\phantom{-}\beta_1&\phantom{-}0\\[0.05cm]
             0&\phantom{-}\beta_1&-\beta_2&\phantom{-}0\\[0.05cm]
             \beta_3&\phantom{-}0&\phantom{-}0&-\beta_0\\[0.05cm]
            \end{smallmatrix}
\right], \;
\mathbf{A}_2=\left[\begin{smallmatrix}
             \ \ &\ \ &\ \ &\ \ \\
             1&\phantom{-}1&\phantom{-}0&\phantom{-}0\\[0.05cm]
             1&-1&\phantom{-}0&\phantom{-}0\\[0.05cm]
             0&\phantom{-}0&-1&\phantom{-}1\\[0.05cm]
             0&\phantom{-}0&\phantom{-}1&\phantom{-}1\\[0.05cm]
            \end{smallmatrix}
\right],\;
\mathbf{A}_3=\left[\begin{smallmatrix}
             \ \ &\ \ &\ \ &\ \ \\
             0&\phantom{-}0&\phantom{-}0&\phantom{-}1\\[0.05cm]
             0&\phantom{-}\alpha&\phantom{-}\alpha&\phantom{-}0\\[0.05cm]
             0&-\alpha&\phantom{-}\alpha&\phantom{-}0\\[0.05cm]
             1&\phantom{-}0&\phantom{-}0&\phantom{-}0\\[0.05cm]
            \end{smallmatrix}
\right],
\end{align*}
with $\alpha=\cos\!\left(\frac{\pi}{4}\right)$ and $\beta_n=\cos\!\left(\frac{(2n+1)\pi}{16}\right)$.

Replacing (\ref{ec:ecuation2}) in (\ref{ec:ecuation1}) and expanding the factorization, we obtain:
\begin{equation}
\label{ec:ecuation3}
\mathbf{C}_8^{\mbox{\scriptsize II}}=\frac{1}{2}\,\mathbf{P}_8\,\mathbf{M}_1\,\mathbf{M}_2\,\mathbf{M}_3\,\mathbf{M}_4\,\mathbf{B}_8\ ,
\end{equation}
where
\begin{align*}
\mathbf{M}_1=&\left[\begin{array}{cc}
                                                         \mathbf{I}_4&\mathbf{0}_4 \\
                                                         \mathbf{0}_4& \bar{\mathbf{I}}_4\,\mathbf{Q}
                                                        \end{array}\right],\ \mathbf{M}_2=\left[\begin{array}{cc}
                                                         \mathbf{P}_4& \mathbf{0}_4\\
                                                         \mathbf{0}_4& \mathbf{A}_1
                                                        \end{array}\right],\ \mathbf{M}_3=\left[\begin{array}{cc}
                                                         \widetilde{\mathbf{C}}& \mathbf{0}_4\\
                                                         \mathbf{0}_4& \mathbf{A}_2
                                                        \end{array}\right], \\
\mathbf{M}_4=&\left[\begin{array}{cc}
                                                         \mathbf{B}_4&\mathbf{0}_4 \\
                                                         \mathbf{0}_4& \mathbf{A}_3
                                                        \end{array}\right],\ \mathbf{P}_4=\left[\begin{smallmatrix}
             \ \ &\ \ &\ \ &\ \ \\
             1&0&0&0\\[0.05cm]
             0&0&0&1\\[0.05cm]
             0&1&0&0\\[0.05cm]
             0&0&1&0\\[0.05cm]
            \end{smallmatrix}
\right],\ \mathbf{B}_4=\left[\begin{array}{rr}
           \mathbf{I}_2&\bar{\mathbf{I}}_2\\
           \bar{\mathbf{I}}_2&-\mathbf{I}_2
          \end{array}
\right]
, \\
\widetilde{\mathbf{C}}=&\left[\begin{array}{cc}    \mathbf{C}_2^{\mbox{\scriptsize II}} & \mathbf{0}_4\\
\mathbf{0}_4& \bar{\mathbf{I}}_2\,\mathbf{C}_2^{\mbox{\scriptsize IV}}\,\bar{\mathbf{I}}_2
\end{array}\right]=\left[\begin{smallmatrix}
             \ \ &\ \ &\ \ &\ \ \\
             \alpha&\phantom{-}\alpha&\phantom{-}0&\phantom{-}0\\[0.05cm]
             \alpha&-\alpha&\phantom{-}0&\phantom{-}0\\[0.05cm]
             0&\phantom{-}0&-\gamma_0&\phantom{-}\gamma_1\\[0.05cm]
             0&\phantom{-}0&\phantom{-}\gamma_1&\phantom{-}\gamma_0\\[0.05cm]
            \end{smallmatrix}
\right],
\end{align*}
with
$\gamma_n=\cos\!\left(\frac{(2n+1)\pi}{8}\right)$.
The expression in~\eqref{ec:ecuation3}
is
referred to as
Chen's factorization
for the 8-point DCT-II.

Without any fast algorithm,
the computation of the DCT-II
requires 64~multiplications and 56~additions.
Using the Chen's factorization in~\eqref{ec:ecuation3}
the arithmetic complexity is reduced to
16~multiplications and 26~additions.
The quantities
$\alpha$, $\beta_n$, and $\gamma_n$,
presented in
$\mathbf{M}_2$, $\mathbf{M}_3$, and $\mathbf{M}_4$,
are
irrational numbers
and demand non-trivial multiplications.

For the sake of notation,
hereafter the DCT-II is referred to as DCT.

\section{Proposed DCT approximations}~\label{s:DCT_approximations}

In this section,
new approximations for the DCT
are sought.
To this end,
we
notice that the factorization~\eqref{ec:ecuation3}
naturally induces the following mapping:
\begin{align}
\label{eq:Chen_factorization}
 \operatorname{T}_C: \mathbb{R}\times\mathbb{R}^4\times\mathbb{R}^2&\longrightarrow\mathcal{M}_8(\mathbb{R})\nonumber\\
 (\alpha,\boldsymbol{\beta},\boldsymbol{\gamma})&\longrightarrow\mathbf{P}_8\,\mathbf{M}_1\,\mathbf{M}_2\,\mathbf{M}_3\,\mathbf{M}_4\,\mathbf{B}_8,
\end{align}
where
$\mathcal{M}_8(\mathbb{R})$
is the space of 8$\times$8 matrices with real-valued entries,
$\alpha\in\mathbb{R}$,
$\boldsymbol{\beta}=[\beta_0\ \ \beta_1\ \ \beta_2\ \ \beta_3]^{\top}\in\mathbb{R}^4$,
and
$\boldsymbol{\gamma}=[\gamma_0 \ \ \gamma_1]^{\top}\in\mathbb{R}^2$.
Now
the matrices
$\mathbf{M}_2$,
$\mathbf{M}_3$,
and
$\mathbf{M}_4$
are seen as matrix functions,
where the constants in
\eqref{ec:ecuation3}
are understood
as
parameters:
\begin{equation}
\label{equation-free-parameter-matrices}
\begin{split}
\mathbf{M}_2&=\mathbf{M}_2(\boldsymbol{\beta})
,
\\
\mathbf{M}_3&=\mathbf{M}_3(\alpha,\boldsymbol{\gamma})
,
\\
\mathbf{M}_4&=\mathbf{M}_4(\alpha)
.
\end{split}
\end{equation}
In particular,
for the values
\begin{align}
\label{eq:set_values_Chen}
 \begin{split}
  \alpha=\,&\cos\left(\frac{\pi}{4}\right),\\
  \beta_n=\,&\cos\left(\frac{(2n+1)\pi}{16}\right),\ n=0,1,2,3,\\
  \gamma_n=\,&\cos\left(\frac{(2n+1)\pi}{8}\right),\ n=0,1,
 \end{split}
\end{align}
we have
$\operatorname{T}_C(\alpha,\boldsymbol{\beta},\boldsymbol{\gamma})=2\,\mathbf{C}_8^{\mbox{\scriptsize II}}$.
In the following,
we vary the
values of parameters
$\alpha$,
$\boldsymbol{\beta}$,
and
$\boldsymbol{\gamma}$
aiming at the derivation
of low-complexity matrices whose elements are
restricted to the set
$\mathcal{P}=\{0,\pm\frac{1}{2},\pm 1,\pm 2\}$.

To facilitate our approach,
we consider
the signum
and
round-off functions,
respectively,
given by:
\begin{align*}
\operatorname{sign}(x)
&=
\begin{cases}
                         1,&\ \mbox{if}\ x>0,\\
                         0,&\ \mbox{if}\ x=0,\\
                         -1,&\ \mbox{if}\ x<0,
                \end{cases}
\\
\operatorname{round}(x)
&=
\operatorname{sign}(x) \, \left\lfloor \left\vert x \right\vert + \frac{1}{2}\right\rfloor
\end{align*}
where
$ \lfloor x \rfloor
=
\max\, \{m\in\mathbb{Z}\mid m\le x\}$
is the floor function
for $x\in\mathbb{R}$.
These functions coincide with their definitions
implemented in C and {\small\textsc{MATLAB}} computer languages.
When applied to vectors or matrices,
$\operatorname{sign}(\cdot)$ and $\operatorname{round}(\cdot)$
operate entry-wise.

Thereby, considering the values in~\eqref{eq:set_values_Chen} and
applying directly the functions above,
we obtain the approximate vectors shown below:
\begin{align*}
 (\widetilde{\alpha},\widetilde{\boldsymbol{\beta}},\widetilde{\boldsymbol{\gamma}})=&\operatorname{sign}[(\alpha,\boldsymbol{\beta},\boldsymbol{\gamma})]=\left[1\ \ 1\ \ 1\ \ 1\ \ 1\ \ 1 \ \ 1\right]^{\top},\\
 (\widehat{\alpha},\widehat{\boldsymbol{\beta}},\widehat{\boldsymbol{\gamma}})=&\operatorname{round}[(\alpha,\boldsymbol{\beta},\boldsymbol{\gamma})]=\left[1\ \ 1\ \ 0\ \ 1\ \ 1\ \ 1 \ \ 0\right]^{\top}.
\end{align*}
Then,
the following matrices are generated
according to~\eqref{equation-free-parameter-matrices}:
\begin{align*}
\widetilde{\mathbf{M}}_2=\,&\mathbf{M}_2(\widetilde{\boldsymbol{\beta}}),
\ \widehat{\mathbf{M}}_2=\mathbf{M}_2(\widehat{\boldsymbol{\beta}}),\\
\widetilde{\mathbf{M}}_3=\,&\mathbf{M}_3(\widetilde{\alpha},\widetilde{\boldsymbol{\gamma}}),
\ \widehat{\mathbf{M}}_3=\mathbf{M}_3(\widehat{\alpha},\widehat{\boldsymbol{\gamma}}),\\
\widetilde{\mathbf{M}}_4=\,&\mathbf{M}_4(\widetilde{\alpha}),
\ \widehat{\mathbf{M}}_4=\mathbf{M}_4(\widehat{\alpha})
,
\end{align*}
which are explicitly given by:
\begin{align*}
\widetilde{\mathbf{M}}_2=&\begin{bmatrix}
			    \begin{smallmatrix}
			    &&&&&&&\\
			     1&\phantom{-}0&\phantom{-}0&\phantom{-}0&\phantom{-}0&\phantom{-}0&\phantom{-}0&\phantom{-}0\\
			     0&\phantom{-}0&\phantom{-}0&\phantom{-}1&\phantom{-}0&\phantom{-}0&\phantom{-}0&\phantom{-}0\\
			     0&\phantom{-}1&\phantom{-}0&\phantom{-}0&\phantom{-}0&\phantom{-}0&\phantom{-}0&\phantom{-}0\\
			     0&\phantom{-}0&\phantom{-}1&\phantom{-}0&\phantom{-}0&\phantom{-}0&\phantom{-}0&\phantom{-}0\\
			     0&\phantom{-}0&\phantom{-}0&\phantom{-}0&\phantom{-}1&\phantom{-}0&\phantom{-}0&\phantom{-}1\\
			     0&\phantom{-}0&\phantom{-}0&\phantom{-}0&\phantom{-}0&\phantom{-}1&\phantom{-}1&\phantom{-}0\\
			     0&\phantom{-}0&\phantom{-}0&\phantom{-}0&\phantom{-}0&\phantom{-}1&-1&\phantom{-}0\\
			     0&\phantom{-}0&\phantom{-}0&\phantom{-}0&\phantom{-}1&\phantom{-}0&\phantom{-}0&-1\\[0.05cm]
			    \end{smallmatrix}
                          \end{bmatrix}, \;
                          \ \ \widehat{\mathbf{M}}_2=\begin{bmatrix}
			    \begin{smallmatrix}
			    &&&&&&&\\
			     1&\phantom{-}0&\phantom{-}0&\phantom{-}0&\phantom{-}0&\phantom{-}0&\phantom{-}0&\phantom{-}0\\
			     0&\phantom{-}0&\phantom{-}0&\phantom{-}1&\phantom{-}0&\phantom{-}0&\phantom{-}0&\phantom{-}0\\
			     0&\phantom{-}1&\phantom{-}0&\phantom{-}0&\phantom{-}0&\phantom{-}0&\phantom{-}0&\phantom{-}0\\
			     0&\phantom{-}0&\phantom{-}1&\phantom{-}0&\phantom{-}0&\phantom{-}0&\phantom{-}0&\phantom{-}0\\
			     0&\phantom{-}0&\phantom{-}0&\phantom{-}0&\phantom{-}1&\phantom{-}0&\phantom{-}0&\phantom{-}0\\
			     0&\phantom{-}0&\phantom{-}0&\phantom{-}0&\phantom{-}0&\phantom{-}1&\phantom{-}1&\phantom{-}0\\
			     0&\phantom{-}0&\phantom{-}0&\phantom{-}0&\phantom{-}0&\phantom{-}1&-1&\phantom{-}0\\
			     0&\phantom{-}0&\phantom{-}0&\phantom{-}0&\phantom{-}0&\phantom{-}0&\phantom{-}0&-1\\[0.05cm]
			    \end{smallmatrix}
			   \end{bmatrix},\\
			   \widetilde{\mathbf{M}}_3=&\begin{bmatrix}
			    \begin{smallmatrix}
			    &&&&&&&\\
			     1&\phantom{-}1&\phantom{-}0&\phantom{-}0&\phantom{-}0&\phantom{-}0&\phantom{-}0&\phantom{-}0\\
			     1&-1&\phantom{-}0&\phantom{-}0&\phantom{-}0&\phantom{-}0&\phantom{-}0&\phantom{-}0\\
			     0&\phantom{-}0&-1&\phantom{-}1&\phantom{-}0&\phantom{-}0&\phantom{-}0&\phantom{-}0\\
			     0&\phantom{-}0&\phantom{-}1&\phantom{-}1&\phantom{-}0&\phantom{-}0&\phantom{-}0&\phantom{-}0\\
			     0&\phantom{-}0&\phantom{-}0&\phantom{-}0&\phantom{-}1&\phantom{-}1&\phantom{-}0&\phantom{-}0\\
			     0&\phantom{-}0&\phantom{-}0&\phantom{-}0&\phantom{-}1&-1&\phantom{-}0&\phantom{-}0\\
			     0&\phantom{-}0&\phantom{-}0&\phantom{-}0&\phantom{-}0&\phantom{-}0&-1&\phantom{-}1\\
			     0&\phantom{-}0&\phantom{-}0&\phantom{-}0&\phantom{-}0&\phantom{-}0&\phantom{-}1&\phantom{-}1\\[0.05cm]
			    \end{smallmatrix}
                          \end{bmatrix},    \;
                          \ \ \widehat{\mathbf{M}}_3=\begin{bmatrix}
			    \begin{smallmatrix}
			    &&&&&&&\\
			     1&\phantom{-}1&\phantom{-}0&\phantom{-}0&\phantom{-}0&\phantom{-}0&\phantom{-}0&\phantom{-}0\\
			     1&-1&\phantom{-}0&\phantom{-}0&\phantom{-}0&\phantom{-}0&\phantom{-}0&\phantom{-}0\\
			     0&\phantom{-}0&-1&\phantom{-}0&\phantom{-}0&\phantom{-}0&\phantom{-}0&\phantom{-}0\\
			     0&\phantom{-}0&\phantom{-}0&\phantom{-}1&\phantom{-}0&\phantom{-}0&\phantom{-}0&\phantom{-}0\\
			     0&\phantom{-}0&\phantom{-}0&\phantom{-}0&\phantom{-}1&\phantom{-}1&\phantom{-}0&\phantom{-}0\\
			     0&\phantom{-}0&\phantom{-}0&\phantom{-}0&\phantom{-}1&-1&\phantom{-}0&\phantom{-}0\\
			     0&\phantom{-}0&\phantom{-}0&\phantom{-}0&\phantom{-}0&\phantom{-}0&-1&\phantom{-}1\\
			     0&\phantom{-}0&\phantom{-}0&\phantom{-}0&\phantom{-}0&\phantom{-}0&\phantom{-}1&\phantom{-}1\\[0.05cm]
			    \end{smallmatrix}
                          \end{bmatrix},\\
                          \widetilde{\mathbf{M}}_4=&\begin{bmatrix}
			    \begin{smallmatrix}
			    &&&&&&&\\
			     1&\phantom{-}0&\phantom{-}0&\phantom{-}1&\phantom{-}0&\phantom{-}0&\phantom{-}0&\phantom{-}0\\
			     0&\phantom{-}1&\phantom{-}1&\phantom{-}0&\phantom{-}0&\phantom{-}0&\phantom{-}0&\phantom{-}0\\
			     0&\phantom{-}1&-1&\phantom{-}0&\phantom{-}0&\phantom{-}0&\phantom{-}0&\phantom{-}0\\
			     1&\phantom{-}0&\phantom{-}0&-1&\phantom{-}0&\phantom{-}0&\phantom{-}0&\phantom{-}0\\
			     0&\phantom{-}0&\phantom{-}0&\phantom{-}0&\phantom{-}0&\phantom{-}0&\phantom{-}0&\phantom{-}1\\
			     0&\phantom{-}0&\phantom{-}0&\phantom{-}0&\phantom{-}0&\phantom{-}1&\phantom{-}1&\phantom{-}0\\
			     0&\phantom{-}0&\phantom{-}0&\phantom{-}0&\phantom{-}0&-1&\phantom{-}1&\phantom{-}0\\
			     0&\phantom{-}0&\phantom{-}0&\phantom{-}0&\phantom{-}1&\phantom{-}0&\phantom{-}0&\phantom{-}0\\[0.05cm]
			    \end{smallmatrix}
                          \end{bmatrix}=\widehat{\mathbf{M}}_4.\\
\end{align*}

Invoking the factorization from~\eqref{eq:Chen_factorization},
we define the following new transforms:
\begin{align}
\widetilde{\mathbf{T}}_8
\triangleq
\operatorname{T}_C
\left(
\widetilde{\alpha},\widetilde{\boldsymbol{\beta}},\widetilde{\boldsymbol{\gamma}}
\right)
=&
\,
\mathbf{P}_8\,
\mathbf{M}_1\,
\widetilde{\mathbf{M}}_2\,
\widetilde{\mathbf{M}}_3\,
\widetilde{\mathbf{M}}_4\,
\mathbf{B}_8
,
\label{ec:ecuation4}
\\
\widehat{\mathbf{T}}_8
\triangleq
\operatorname{T}_C
\left(
\widehat{\alpha},\widehat{\boldsymbol{\beta}},\widehat{\boldsymbol{\gamma}}
\right)
=&\,
\mathbf{P}_8\,
\mathbf{M}_1\,
\widehat{\mathbf{M}}_2\,
\widehat{\mathbf{M}}_3\,
\widehat{\mathbf{M}}_4\,
\mathbf{B}_8
.
\label{ec:ecuation5}
\end{align}
The numerical evaluation of
\eqref{ec:ecuation4}
and~\eqref{ec:ecuation5}
reveals
the
following matrix transforms:
\begin{align*}
\widetilde{\mathbf{T}}_8=\left[\begin{smallmatrix}
                   &\ \ \ &\ \ \ &\ \ \ &\ \ \ &\ \ \ &\ \ \ &\ \ \ \\[0.05cm]
                   1&\phantom{-}1&\phantom{-}1&\phantom{-}1&\phantom{-}1&\phantom{-}1&\phantom{-}1&\phantom{-}1\\[0.05cm]
                   1&\phantom{-}2&\phantom{-}0&\phantom{-}1&-1&\phantom{-}0&-2&-1\\[0.05cm]
                   1&\phantom{-}1&-1&-1&-1&-1&\phantom{-}1&\phantom{-}1\\[0.05cm]
                   1&\phantom{-}0&-2&-1&\phantom{-}1&\phantom{-}2&\phantom{-}0&-1\\[0.05cm]
                   1&-1&-1&\phantom{-}1&\phantom{-}1&-1&-1&\phantom{-}1\\[0.05cm]
                   1&-2&\phantom{-}0&\phantom{-}1&-1&\phantom{-}0&\phantom{-}2&-1\\[0.05cm]
                   1&-1&\phantom{-}1&-1&-1&\phantom{-}1&-1&\phantom{-}1\\[0.05cm]
                   1&\phantom{-}0&\phantom{-}2&-1&\phantom{-}1&-2&\phantom{-}0&-1\\[0.05cm]
                  \end{smallmatrix}
\right],\;
\widehat{\mathbf{T}}_8=\left[\begin{smallmatrix}
                   &\ \ \ &\ \ \ &\ \ \ &\ \ \ &\ \ \ &\ \ \ &\ \ \ \\[0.05cm]
                   1&\phantom{-}1&\phantom{-}1&\phantom{-}1&\phantom{-}1&\phantom{-}1&\phantom{-}1&\phantom{-}1\\[0.05cm]
                   1&\phantom{-}1&\phantom{-}1&\phantom{-}0&\phantom{-}0&-1&-1&-1\\[0.05cm]
                   1&\phantom{-}0&\phantom{-}0&-1&-1&\phantom{-}0&\phantom{-}0&\phantom{-}1\\[0.05cm]
                   1&\phantom{-}0&-2&-1&\phantom{-}1&\phantom{-}2&\phantom{-}0&-1\\[0.05cm]
                   1&-1&-1&\phantom{-}1&\phantom{-}1&-1&-1&\phantom{-}1\\[0.05cm]
                   1&-2&\phantom{-}0&\phantom{-}1&-1&\phantom{-}0&\phantom{-}2&-1\\[0.05cm]
                   0&-1&\phantom{-}1&\phantom{-}0&\phantom{-}0&\phantom{-}1&-1&\phantom{-}0\\[0.05cm]
                   0&-1&\phantom{-}1&-1&\phantom{-}1&-1&\phantom{-}1&\phantom{-}0\\[0.05cm]
                  \end{smallmatrix}
\right].
\end{align*}

Above transformations
have simple inverse matrices.
Direct matrix inversion rules
applied to
\eqref{ec:ecuation4} and \eqref{ec:ecuation5}
furnish:
\begin{align}
\widetilde{\mathbf{T}}_8^{-1}=&\,\frac{1}{2}\,\mathbf{B}_8^\top\,\widetilde{\mathbf{M}}_4^{-1}\,\widetilde{\mathbf{M}}_3^{-1}\,\widetilde{\mathbf{M}}_2^{-1}\,\mathbf{M}_1^\top\,\mathbf{P}_8^\top,\label{ec:ecuation6}\\
 \widehat{\mathbf{T}}_8^{-1}=&\,\frac{1}{2}\,\mathbf{B}_8^\top\,\widehat{\mathbf{M}}_4^{-1}\, \widehat{\mathbf{M}}_3^{-1}\,\widehat{\mathbf{M}}_2^{-1}\,\mathbf{M}_1^\top\,\mathbf{P}_8^\top
,
\label{ec:ecuation7}
\end{align}
where
\begin{align*}
\mathbf{P}_8^{-1}=&\,\mathbf{P}_8^{\top},\;
\mathbf{M}_1^{-1}=\,\mathbf{M}_1^{\top}, \;
\widetilde{\mathbf{M}}_2^{-1}=\,\frac{1}{2}\,\left[\begin{smallmatrix}
                   &&&&&&&\\[0.05cm]
                   2&\phantom{-}0&\phantom{-}0&\phantom{-}0&\phantom{-}0&\phantom{-}0&\phantom{-}0&\phantom{-}0\\[0.05cm]
                   0&\phantom{-}0&\phantom{-}2&\phantom{-}0&\phantom{-}0&\phantom{-}0&\phantom{-}0&\phantom{-}0\\[0.05cm]
                   0&\phantom{-}0&\phantom{-}0&\phantom{-}2&\phantom{-}0&\phantom{-}0&\phantom{-}0&\phantom{-}0\\[0.05cm]
                   0&\phantom{-}2&\phantom{-}0&\phantom{-}0&\phantom{-}0&\phantom{-}0&\phantom{-}0&\phantom{-}0\\[0.05cm]
                   0&\phantom{-}0&\phantom{-}0&\phantom{-}0&\phantom{-}1&\phantom{-}0&\phantom{-}0&\phantom{-}1\\[0.05cm]
                   0&\phantom{-}0&\phantom{-}0&\phantom{-}0&\phantom{-}0&\phantom{-}1&\phantom{-}1&\phantom{-}0\\[0.05cm]
                   0&\phantom{-}0&\phantom{-}0&\phantom{-}0&\phantom{-}0&\phantom{-}1&-1&\phantom{-}0\\[0.05cm]
                   0&\phantom{-}0&\phantom{-}0&\phantom{-}0&\phantom{-}1&\phantom{-}0&\phantom{-}0&-1\\[0.05cm]
                  \end{smallmatrix}
\right],\;
\widehat{\mathbf{M}}_2^{-1}=\frac{1}{2}\,\left[\begin{smallmatrix}
                   &&&&&&&\\[0.05cm]
                   2&\phantom{-}0&\phantom{-}0&\phantom{-}0&\phantom{-}0&\phantom{-}0&\phantom{-}0&\phantom{-}0\\[0.05cm]
                   0&\phantom{-}0&\phantom{-}2&\phantom{-}0&\phantom{-}0&\phantom{-}0&\phantom{-}0&\phantom{-}0\\[0.05cm]
                   0&\phantom{-}0&\phantom{-}0&\phantom{-}2&\phantom{-}0&\phantom{-}0&\phantom{-}0&\phantom{-}0\\[0.05cm]
                   0&\phantom{-}2&\phantom{-}0&\phantom{-}0&\phantom{-}0&\phantom{-}0&\phantom{-}0&\phantom{-}0\\[0.05cm]
                   0&\phantom{-}0&\phantom{-}0&\phantom{-}0&\phantom{-}2&\phantom{-}0&\phantom{-}0&\phantom{-}0\\[0.05cm]
                   0&\phantom{-}0&\phantom{-}0&\phantom{-}0&\phantom{-}0&\phantom{-}1&\phantom{-}1&\phantom{-}0\\[0.05cm]
                   0&\phantom{-}0&\phantom{-}0&\phantom{-}0&\phantom{-}0&\phantom{-}1&-1&\phantom{-}0\\[0.05cm]
                   0&\phantom{-}0&\phantom{-}0&\phantom{-}0&\phantom{-}0&\phantom{-}0&\phantom{-}0&-2\\[0.05cm]
                  \end{smallmatrix}
\right]
,\\
\widetilde{\mathbf{M}}_3^{-1}=&\,\frac{1}{2}\,\widetilde{\mathbf{M}}_3,\;
\widehat{\mathbf{M}}_3^{-1}=\,\frac{1}{2}\,\left[\begin{smallmatrix}
                   &&&&&&&\\[0.05cm]
                   1&\phantom{-}1&\phantom{-}0&\phantom{-}0&\phantom{-}0&\phantom{-}0&\phantom{-}0&\phantom{-}0\\[0.05cm]
                   1&-1&\phantom{-}0&\phantom{-}0&\phantom{-}0&\phantom{-}0&\phantom{-}0&\phantom{-}0\\[0.05cm]
                   0&\phantom{-}0&-2&\phantom{-}0&\phantom{-}0&\phantom{-}0&\phantom{-}0&\phantom{-}0\\[0.05cm]
                   0&\phantom{-}0&\phantom{-}0&2&\phantom{-}0&\phantom{-}0&\phantom{-}0&\phantom{-}0\\[0.05cm]
                   0&\phantom{-}0&\phantom{-}0&\phantom{-}0&\phantom{-}1&\phantom{-}1&\phantom{-}0&\phantom{-}0\\[0.05cm]
                   0&\phantom{-}0&\phantom{-}0&\phantom{-}0&\phantom{-}1&-1&\phantom{-}0&\phantom{-}0\\[0.05cm]
                   0&\phantom{-}0&\phantom{-}0&\phantom{-}0&\phantom{-}0&\phantom{-}0&-1&\phantom{-}1\\[0.05cm]
                   0&\phantom{-}0&\phantom{-}0&\phantom{-}0&\phantom{-}0&\phantom{-}0&\phantom{-}1&\phantom{-}1\\[0.05cm]
                  \end{smallmatrix}
\right],\;
\widetilde{\mathbf{M}}_4^{-1}=\frac{1}{2}\,\left[\begin{smallmatrix}
                   &&&&&&&\\[0.05cm]
                   1&\phantom{-}0&\phantom{-}0&\phantom{-}1&\phantom{-}0&\phantom{-}0&\phantom{-}0&\phantom{-}0\\[0.05cm]
                   0&\phantom{-}1&\phantom{-}1&\phantom{-}0&\phantom{-}0&\phantom{-}0&\phantom{-}0&\phantom{-}0\\[0.05cm]
                   0&\phantom{-}1&-1&\phantom{-}0&\phantom{-}0&\phantom{-}0&\phantom{-}0&\phantom{-}0\\[0.05cm]
                   1&\phantom{-}0&\phantom{-}0&-1&\phantom{-}0&\phantom{-}0&\phantom{-}0&\phantom{-}0\\[0.05cm]
                   0&\phantom{-}0&\phantom{-}0&\phantom{-}0&\phantom{-}0&\phantom{-}0&\phantom{-}0&\phantom{-}2\\[0.05cm]
                   0&\phantom{-}0&\phantom{-}0&\phantom{-}0&\phantom{-}0&\phantom{-}1&-1&\phantom{-}0\\[0.05cm]
                   0&\phantom{-}0&\phantom{-}0&\phantom{-}0&\phantom{-}0&\phantom{-}1&\phantom{-}1&\phantom{-}0\\[0.05cm]
                   0&\phantom{-}0&\phantom{-}0&\phantom{-}0&\phantom{-}2&\phantom{-}0&\phantom{-}0&\phantom{-}0\\[0.05cm]
                  \end{smallmatrix}
\right]=\widehat{\mathbf{M}}_4^{-1}
,
\end{align*}
and
\begin{align*}
\mathbf{B}_8^{-1}=&\,\frac{1}{2}\,\mathbf{B}_8^{\top}.
\end{align*}

\begin{figure}
\centering
\subfigure[$\widehat{\mathbf{T}}_8$]
{\includegraphics[width=0.4\linewidth]{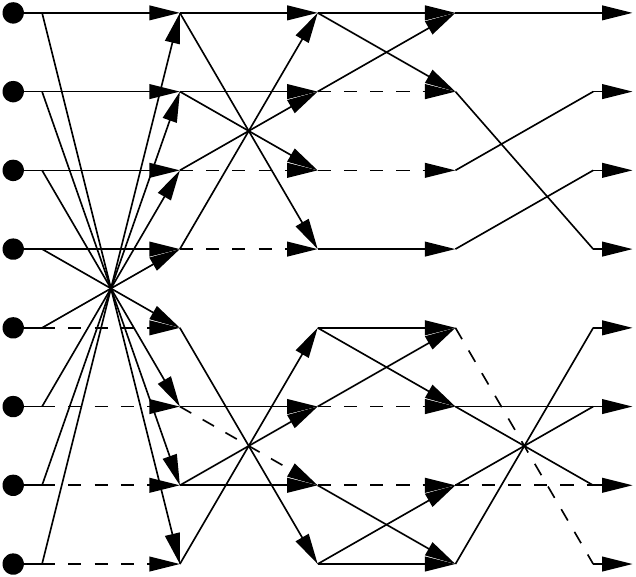}\label{f:direct}
 \put(-198,165){\footnotesize $x_0$}
  \put(-198,142){\footnotesize $x_1$}
  \put(-198,118){\footnotesize $x_2$}
  \put(-198,95){\footnotesize $x_3$}
 \put(-198,72){\footnotesize $x_4$}
 \put(-198,49){\footnotesize $x_5$}
 \put(-198,25){\footnotesize $x_6$}
 \put(-198,2){\footnotesize $x_7$}
 \put(0,165){\footnotesize $X_0$}
 \put(0,142){\footnotesize $X_2$}
 \put(0,118){\footnotesize $X_4$}
 \put(0,95){\footnotesize $X_6$}
 \put(0,72){\footnotesize $X_7$}
 \put(0,49){\footnotesize $X_5$}
 \put(0,25){\footnotesize $X_3$}
  \put(0,2){\footnotesize $X_1$}}\hspace*{1.2cm}
 \subfigure[$\widehat{\mathbf{T}}_8^{-1}$]
{\includegraphics[width=0.4\linewidth]{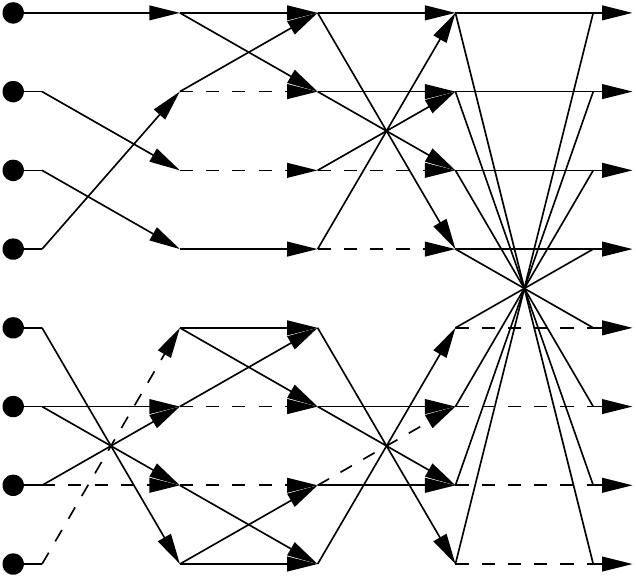}\label{f:inverse}
 \put(-200,165){\footnotesize $X_0$}
  \put(-200,142){\footnotesize $X_2$}
  \put(-200,118){\footnotesize $X_4$}
  \put(-200,95){\footnotesize $X_6$}
 \put(-200,72){\footnotesize $X_7$}
 \put(-200,49){\footnotesize $X_5$}
 \put(-200,25){\footnotesize $X_3$}
 \put(-200,2){\footnotesize $X_1$}
 \put(1,165){\footnotesize $x_0$}
 \put(1,142){\footnotesize $x_1$}
 \put(1,118){\footnotesize $x_2$}
 \put(1,95){\footnotesize $x_3$}
 \put(1,72){\footnotesize $x_4$}
 \put(1,49){\footnotesize $x_5$}
 \put(1,25){\footnotesize $x_6$}
 \put(1,2){\footnotesize $x_7$}
 \put(-181,52){\scriptsize $1/2$}
 \put(-178,40){\scriptsize $1/2$}
 \put(-178,33){\scriptsize $1/2$}
 \put(-181,21){\scriptsize $1/2$}
 \put(-125,122){\scriptsize $-2$}
 \put(-120,99){\scriptsize $2$}
 \put(-95,62){\scriptsize $2$}
 \put(-94,12){\scriptsize $2$}}
\caption[]{\label{figura2}Signal flow graph for $\widehat{\mathbf{T}}_8$ and $\widehat{\mathbf{T}}_8^{-1}$ (dotted lines indicates multiplication by~$-1$).
}
\end{figure}
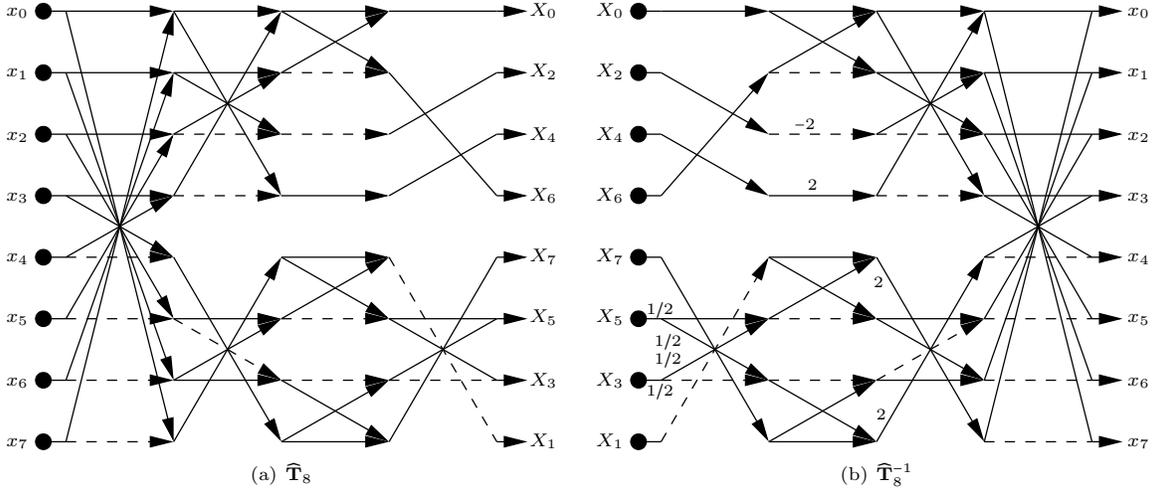

Figure~\ref{figura2} depicts
the signal flow graphs (SFG)
of the proposed fast algorithm
for the transform
$\widehat{\mathbf{T}}_8$ and its inverse $\widehat{\mathbf{T}}_8^{-1}$.
Figure~\ref{f:direct} and \ref{f:inverse} are linked to
\eqref{ec:ecuation5} and \eqref{ec:ecuation7}, respectively.
The SFGs for the
$\widetilde{\mathbf{T}}_8$ and $\widetilde{\mathbf{T}}_8^{-1}$
are similar and were suppressed for brevity.

\subsection{Orthogonality and near orthogonality}
\label{sec: orthogonality}

The proposed transforms
lead
to nonorthogonal approximations for the DCT.
This is also the case
for
the well-known SDCT~\cite{Haweel2001}
and
the BAS approximation described in~\cite{Bouguezel2008}.
Indeed,
for image/video processing,
orthogonality
is not strict requirement;
being near orthogonality
sufficient for very good energy compaction properties.

To quantify how close a given matrix is to an orthogonal matrix,
we adopt the deviation from diagonality measure~\cite{Flury1986},
which is described as follows.
Let~$\mathbf{M}$
be a square matrix.
The deviation from diagonality %
of~$\mathbf{M}$
is given by:
\begin{align*}
\operatorname{\delta}(\mathbf{M})
=
1 -
\frac{\|\operatorname{diag}(\mathbf{M}) \|_\text{F}^2}
{\| \mathbf{M}\|_\text{F}^2}
,
\end{align*}
where $\|\cdot\|_\text{F}$ denotes
the Frobenius norm~\cite[p.~115]{Watkins2004}.
For diagonal matrices, the function $\delta(\cdot)$ returns zero.
Therefore
for a full rank low-complexity transformation matrix $\mathbf{T}$, we can measure
its closeness to orthogonality
by calculating $\operatorname{\delta}(\mathbf{T}\,\mathbf{T}^ \top)$.

Both the 8-point SDCT~\cite{Haweel2001}
and the BAS approximation proposed in~\cite{Bouguezel2008}
are nonorthogonal
and
good DCT approximations.
Their orthogonalization matrices
have
deviation from diagonality
equal to
0.20 and 0.1774,
respectively.
Comparatively,
matrices
$\widetilde{\mathbf{T}}_8$ and $\widehat{\mathbf{T}}_8$
have deviation from diagonality equal to $0.0714$ and $0.0579$, respectively.
Because these values are smaller than those for the SCDT and BAS approximations,
the proposed transformations are more ``nearly orthogonal'' than such approximations.

In~\cite{Tablada2015},
the problem
of deriving DCT approximations
based on nonorthogonal matrices
was addressed.
The approach consists of a variation
of the polar decomposition method as described in~\cite{Higham1986}.
Let
$\mathbf{T}$
be
a full rank low-complexity transformation matrix.
If
$\mathbf{T}$ satisfies the condition:
\begin{align}
\label{eq:ortogonality_condition}
 \mathbf{T}\,\mathbf{T}^\top=\mathbf{D},
\end{align}
where $\mathbf{D}$ is a diagonal matrix,
then
it is possible to derive an orthonormal approximation $\hat{\mathbf{C}}$
linked to $\mathbf{T}$.
This is accomplished by means of the polar decomposition~\cite{Higham1986}:
\begin{align*}
 \hat{\mathbf{C}}=\mathbf{S}\,\mathbf{T},
\end{align*}
where
$\mathbf{S}=\sqrt{\left(\mathbf{T}\,\mathbf{T}^{\top}\right)^{-1}}$
is a positive definite matrix
and
$\sqrt{\cdot}$ denotes the matrix square root operation~\cite{Higham1987}.

From the computational point-of-view,
it is desirable that $\mathbf{S}$ be a diagonal matrix~\cite{Cintra2011a}.
In this case,
the computational complexity of $\hat{\mathbf{C}}$
is the same as that of $\mathbf{T}$,
except for the scaling factors in the diagonal matrix $\mathbf{S}$.
Moreover, in several applications,
such scaling factors can be embedded into
related computation step.
For example,
in JPEG-like compression
the quantization step can absorb
the diagonal elements~\cite{Bayer2012,Bouguezel2011,Ortega2004,Cintra2011}.

Since
the transformations
$\widetilde{\mathbf{T}}_8$ and $\widehat{\mathbf{T}}_8$
do
not
satisfy~\eqref{eq:ortogonality_condition},
one may consider approximating $\mathbf{S}$ itself by replacing the off-diagonal elements of $\mathbf{D}$ by zeros, at the expense of not obtaining a precisely orthogonal approximation $\hat{\mathbf{C}}$.
Therefore,
we consider the following approximate matrix for $\mathbf{S}$:
\begin{align}
\label{eq:quase-ortogonality_condition}
\mathbf{S}'=\sqrt{\left[\operatorname{diag}(\mathbf{T}\,\mathbf{T}^\top)\right]^{-1}},
\end{align}
where $\operatorname{diag}(\cdot)$ returns the diagonal matrix from its matrix argument.
Thus,
the near orthogonal approximations for the DCT-II
associated to the proposed transforms
are given by:
\begin{align*}
\widetilde{\mathbf{C}}_8= \widetilde{\mathbf{S}}_8\,\widetilde{\mathbf{T}}_8,\\
\widehat{\mathbf{C}}_8= \widehat{\mathbf{S}}_8\,\widehat{\mathbf{T}}_8,
\end{align*}
where
{\fontsize{9pt}{10pt}
\begin{align*}
 \widetilde{\mathbf{S}}_8=&\,\operatorname{diag}\left(\frac{1}{\sqrt{8}}\,,\,\frac{1}{\sqrt{12}}\,,\,\frac{1}{\sqrt{8}}\,,\,\frac{1}{\sqrt{12}}\,,\,\frac{1}{\sqrt{8}}\,,\,\frac{1}{\sqrt{12}}\,,\,\frac{1}{\sqrt{8}}\,,\,\frac{1}{\sqrt{12}}\right),\\
 \widehat{\mathbf{S}}_8=&\,\operatorname{diag}\left(\frac{1}{\sqrt{8}}\,,\,\frac{1}{\sqrt{6}}\,,\,\frac{1}{2}\,,\,\frac{1}{\sqrt{12}}\,,\,\frac{1}{\sqrt{8}}\,,\,\frac{1}{\sqrt{12}}\,,\,\frac{1}{2}\,,\,\frac{1}{\sqrt{6}}\right).
\end{align*}
}
Notice that $\widetilde{\mathbf{S}}_8$ and $\widehat{\mathbf{S}}_8$
derive from~\eqref{eq:quase-ortogonality_condition}.

\section{Performance assessment and computational cost}~\label{s:assessment}

To measure the proximity
of the
new multiplierless transforms
with respect to the exact DCT,
we elected the
total error energy~\cite{Cintra2011}
as figure of merit.
We also considered
the coding gain relative to the KLT~\cite{Han2013} as
the measure for coding performance evaluation.
For comparison,
we separated
the classical approximations
SDCT~\cite{Haweel2001} and BAS~\cite{Bouguezel2008}---which
are nonorthogonal---as well as the
HT~\cite{Seberry2005}
and
the WHT~\cite{Horadam2010},
both orthogonal.

\subsection{Performance measures}

\subsubsection{Total error energy}

The total error energy
$\epsilon_{\scriptsize\mbox{\,total}}$
is a measure of spectral similarity between the exact DCT and the considered approximate DCT~\cite{Cintra2011}.
Although originally defined over the spectral domain~\cite{Haweel2001},
the total error energy for a given DCT approximation matrix  $\hat{\mathbf{C}}$ can be written as:
\begin{align*}
\epsilon_{\scriptsize\mbox{\,total}}
=
\pi\,\|\mathbf{C}-\hat{\mathbf{C}}\|_{_{F}}^2
,
\end{align*}
where
$\mathbf{C}$ is the exact DCT matrix
and
$\|\cdot\|_{_F}$ denotes the Frobenius norm~\cite{Watkins2004}.
The
total error energy measurements
for all discussed approximations
are listed in Table~\ref{t:error_energy}.

\begin{table}%
\caption{Total error energy of the considered transforms}
\label{t:error_energy}
\begin{center}
\begin{tabular}{ccccccc}
\toprule
&$\widehat{\mathbf{C}}_8$&$\widetilde{\mathbf{C}}_8$&SDCT&BAS&WHT&HT\\
\midrule
$\epsilon_{\scriptsize\mbox{\,total}}$
&$1.79$&$3.64$&$3.32$&$4.12$&$5.05$&$47.61$\\
\bottomrule
\end{tabular}
\end{center}
\end{table}

The proposed approximation $\widehat{\mathbf{C}}_8$
presents
the lower total error energy among all considered transforms,
at the same time that
requires only 22~additions.
The BAS transform,
which possess the smallest arithmetic cost among the considered methods,
presents
a considerably
higher total error energy.
Thus,
$\widehat{\mathbf{C}}_8$ and SDCT
outperform BAS.
Comparatively,
HT and WHT are less suitable approximations for the DCT.

\subsubsection{Coding gain relative to the KLT}

For coding performance evaluation,
images are assumed to be
modeled after
a first-order Markov process
with
correlation coefficient $\rho$,
where $0\leq\rho<1$~\cite{Liang2001,Britanak2010,Han2013}.
Natural images satisfy
the above assumptions~\cite{Liang2001}.
Then,
the
$(m,n)$-th element of
the covariance matrix~$\mathbf{R}_\mathbf{x}$
of the input signal $\mathbf{x}$
is given by
$r^{(\mathbf{x})}_{m,n} = \rho^{|m-n|}$~\cite{Britanak2010}.

Let $\mathbf{h}_k$ and $\mathbf{g}_k$ be the $k$th rows
of
$\hat{\mathbf{C}}$ and $\hat{\mathbf{C}}^{-1}$,
respectively.
Thus,
the coding gain of~$\hat{\mathbf{C}}$
is given by:
\begin{align*}
C_g(\hat{\mathbf{C}})
=10\,\log_{10}\!\left[
\prod_{k=1}^N
\frac{1}{(A_k \, B_k)^{1/N}}\right]
\quad
(\text{in dB})
,
\end{align*}
where
$
A_k
=
\operatorname{su}
[(
\mathbf{h}_k^\top
\,
\mathbf{h}_k
)
\circ
\mathbf{R}_\mathbf{x}]
$,
$\operatorname{su}(\cdot)$
returns the sum of the elements of its matrix
argument~\cite{Merikoski1984},
operator $\circ$ is the element-wise matrix product~\cite{Seber2008},
$
B_k = \| \mathbf{g}_k \|_2^2
$,
and
$\|\cdot\|_2$ is the Euclidean norm.
For orthogonal transforms,
the unified coding gain
collapses
into the usual coding gain as defined in~\cite{Britanak2010,Han2013}.

High coding gain
values
indicate
better energy compaction capability into
the transform domain~\cite{Liang2001}.
In this sense, the KLT is optimal~\cite{Britanak2010,Effros2004}.
Thus,
an appropriate measure for evaluating
the coding gain is given by~\cite{Han2013}:
\begin{align*}
C_g(\hat{\mathbf{C}}) - C_g({\mbox{KLT}}),
\end{align*}
where $C_g({\mbox{KLT}})$ denotes the coding gain corresponding to the KLT.
For example, for $N=8$ and $\rho=0.95$,
the coding gains
linked to the KLT and DCT
are
8{.}8462~dB and 8{.}8259~dB, respectively~\cite{Britanak2010}.
Hence,
the DCT coding gain relative to the KLT is
$C_g({\mathbf{C}}) - C_g({\mbox{KLT}})=-0.0203$.

\begin{figure}
\centering
\includegraphics[width=0.6\linewidth]{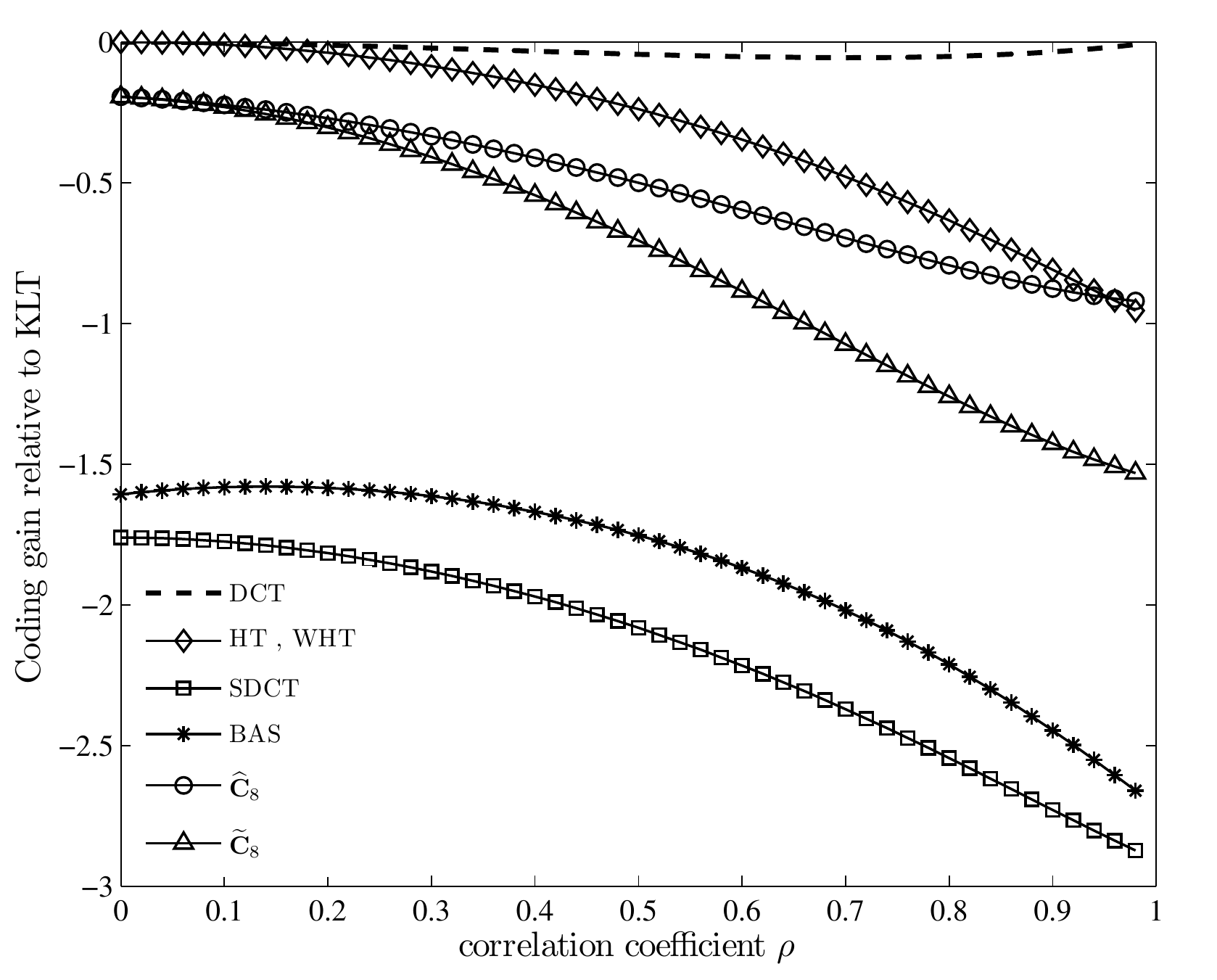}
\caption{Coding gain relative to the KLT
for the considered transforms.}\label{fig:cg}
\end{figure}

Figure~\ref{fig:cg} compares the values of coding gain
relative to the KLT for
the considered transforms
in the range $0\leq\rho<1$.
As expected,
the DCT has the smallest difference with respect to the KLT, followed by
the HT and WHT (both with the same values).
Roughly
orthogonal transforms tend to show
better coding gain performance when compared to nonorthogonal.
The proposed transforms $\widehat{\mathbf{C}}_8$ and $\widetilde{\mathbf{C}}_8$
could outperform the SDCT and BAS, both nonorthogonal transformations.
As $\rho \to 1$,
the approximation
$\widehat{\mathbf{C}}_8$
performs as well as the HT and WHT.
This scenario is realistic
for image compression,
because natural images exhibit
high inter-pixel correlation~\cite{Gonzalez2006}.

\subsection{Computational cost}

The low-complexity matrices
associated to the proposed approximations and their inverses
possess
multiplierless
matrix factorizations
as shown in~\eqref{ec:ecuation4},
\eqref{ec:ecuation5},
\eqref{ec:ecuation6},
and
\eqref{ec:ecuation7}.
Therefore,
the only truly multiplicative elements
are the ones found
in the
diagonal matrices
$\widetilde{\mathbf{S}}_8$
and
$\widehat{\mathbf{S}}_8$.
However,
in the context of image and video coding,
such diagonal matrices
can easily absorbed
in the quantization step~\cite{Bayer2012,Bouguezel2011,Ortega2004,Cintra2011}.
As a consequence,
in that context,
the introduced approximations
can be understood
as fully multiplierless operations.

Table~\ref{t:computational_cost} presents the arithmetic complexity of
the considered transforms.
The computational cost of
the exact DCT
according
to the Chen's factorization
(cf.~\eqref{ec:ecuation3})~\cite{Chen1977}
and the integer DCT for HEVC~\cite{Meher2014}
are also included for comparison.
The proposed approximation $\widehat{\mathbf{C}}_8$
requires only 22~additions.
On the other hand,
the computational cost of
$\widetilde{\mathbf{C}}_8$
is comparatively larger.

\begin{table}[t]%
\caption{Arithmetic complexity of the considered 8-point transforms}
\label{t:computational_cost}
\begin{center}
\begin{tabular}{lcccc}
\toprule
Transform  & Mult & Add  & Shift &Total \\
\midrule%
DCT~\cite{Chen1977} &$16$ &$26$ &$0$&$42$\\
HEVC~\cite{Meher2014} & $0$ & $50$ & $30$ & $80$ \\
$\widehat{\mathbf{C}}_8$ &$0$ &$22$ &$0$&$22$\\
$\widetilde{\mathbf{C}}_8$ &$0$ &$26$ &$0$&$26$\\
SDCT~\cite{Haweel2001} &$0$ &$24$ &$0$&$24$\\
BAS~\cite{Bouguezel2008} &$0$ &$21$ &$0$&$21$\\
WHT~\cite{Horadam2010} &$0$ &$24$ &$0$&$24$\\
HT~\cite{Seberry2005} &$0$ &$24$ &$0$&$24$\\
\bottomrule
\end{tabular}
\end{center}
\end{table}

\section{Low-complexity image compression}
\label{s:image_compression}

In this section,
we describe a JPEG-like image compression computational
experiment.
Proposed transformations are evaluated
in terms of their
data compression capability.

\subsection{JPEG-like compression}

We considered
a set of 30~standard 8-bit 512$\times$512 gray scale images
obtained from~\cite{database2011}.
The images were
subdivided
into
8$\times$8 blocks.
Each block $\mathbf{A}$
was submitted to
the \mbox{2-D} transformation given by~\cite{Suzuki2010}:
\begin{align*}
\hat{\mathbf{B}}=\hat{\mathbf{C}}\, \mathbf{A}\,\hat{\mathbf{C}}^{-1},
\end{align*}
where $\hat{\mathbf{C}}$
is a particular transformation.
The 64~obtained coefficients
in $\hat{\mathbf{B}}$
were arranged
according to the standard zig-zag sequence~\cite{Wallace1992}.
The first $r$~coefficients were retained,
being the remaining coefficients discarded.
We adopted $1\leq r\leq 45$.
Then
for each block $\hat{\mathbf{B}}$,
the \mbox{2-D} inverse transform
was applied
to
reconstruct the compressed image:
\begin{align*}
\hat{\mathbf{A}}=\hat{\mathbf{C}}^{-1}\, \hat{\mathbf{B}}\,\hat{\mathbf{C}}.
\end{align*}
Finally,
the resulting
image
was compared
with the original image
according to the
peak signal-to-noise ratio
(PSNR)~\cite{Huynh-Thu2008}
and
the structural similarity index (SSIM)~\cite{Wang2004, Wang2009}.
The absolute percentage error (APE) for such metrics relative to
the exact DCT was also considered.
The PSNR
is the most commonly used measure for image quality evaluation.
However,
the SSIM measure consider visual human system characteristics
which are not considered by PSNR~\cite{Wang2004}.
Following the methodology proposed in~\cite{Cintra2011},
the average values of the measures over the 30 standard images were computed.
It leads to
statistically
more robust
results
when
compared with
single
image
analysis~\cite{Kay1993}.

\subsection{Results}

\begin{figure}[!t]
\centering
\subfigure[Average PSNR]
{\includegraphics[width=0.49\linewidth]{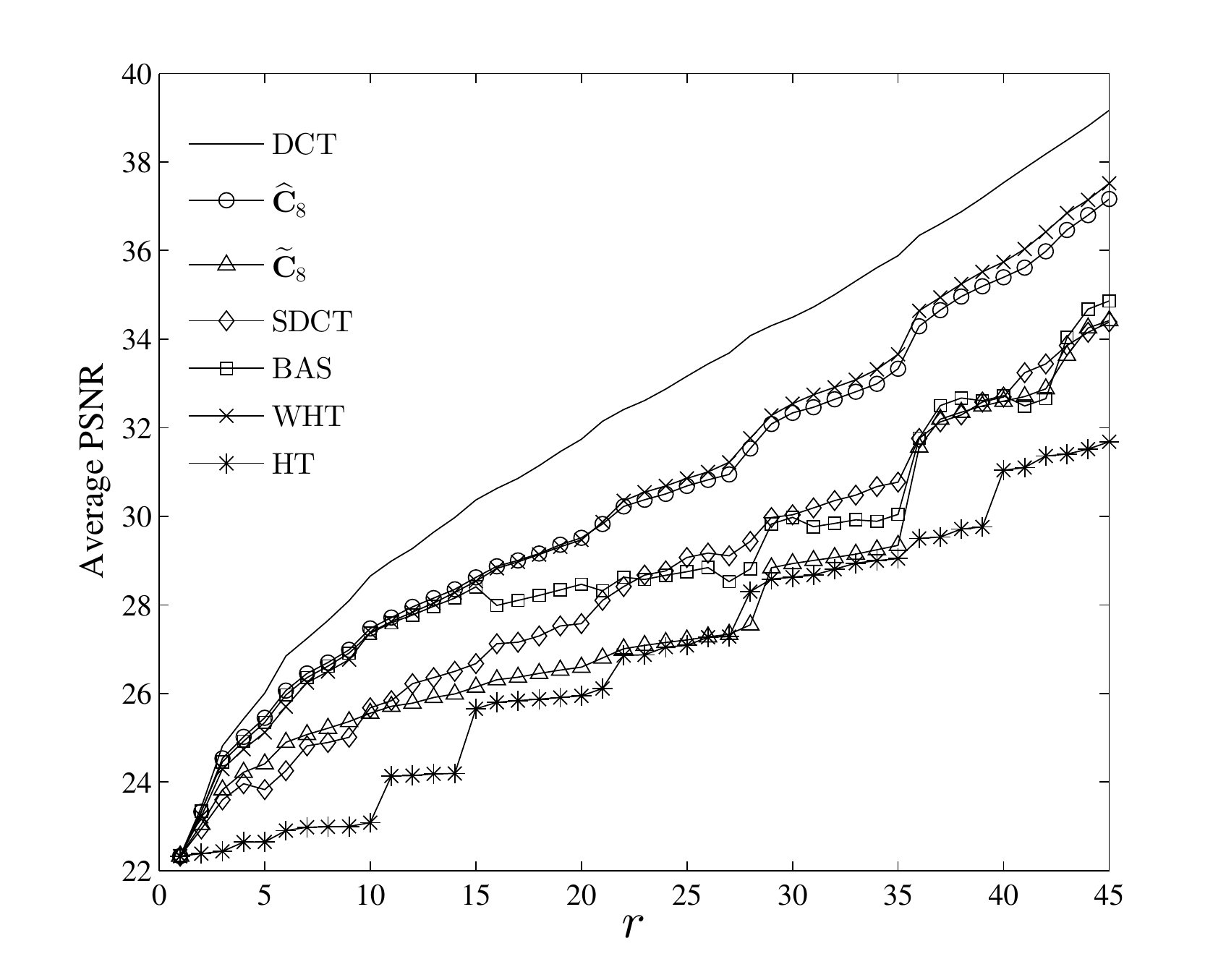}\label{fig:subfig3.1}}
\subfigure[Average PSNR absolute percentage error relative to the DCT.
]
{\includegraphics[width=0.49\linewidth]{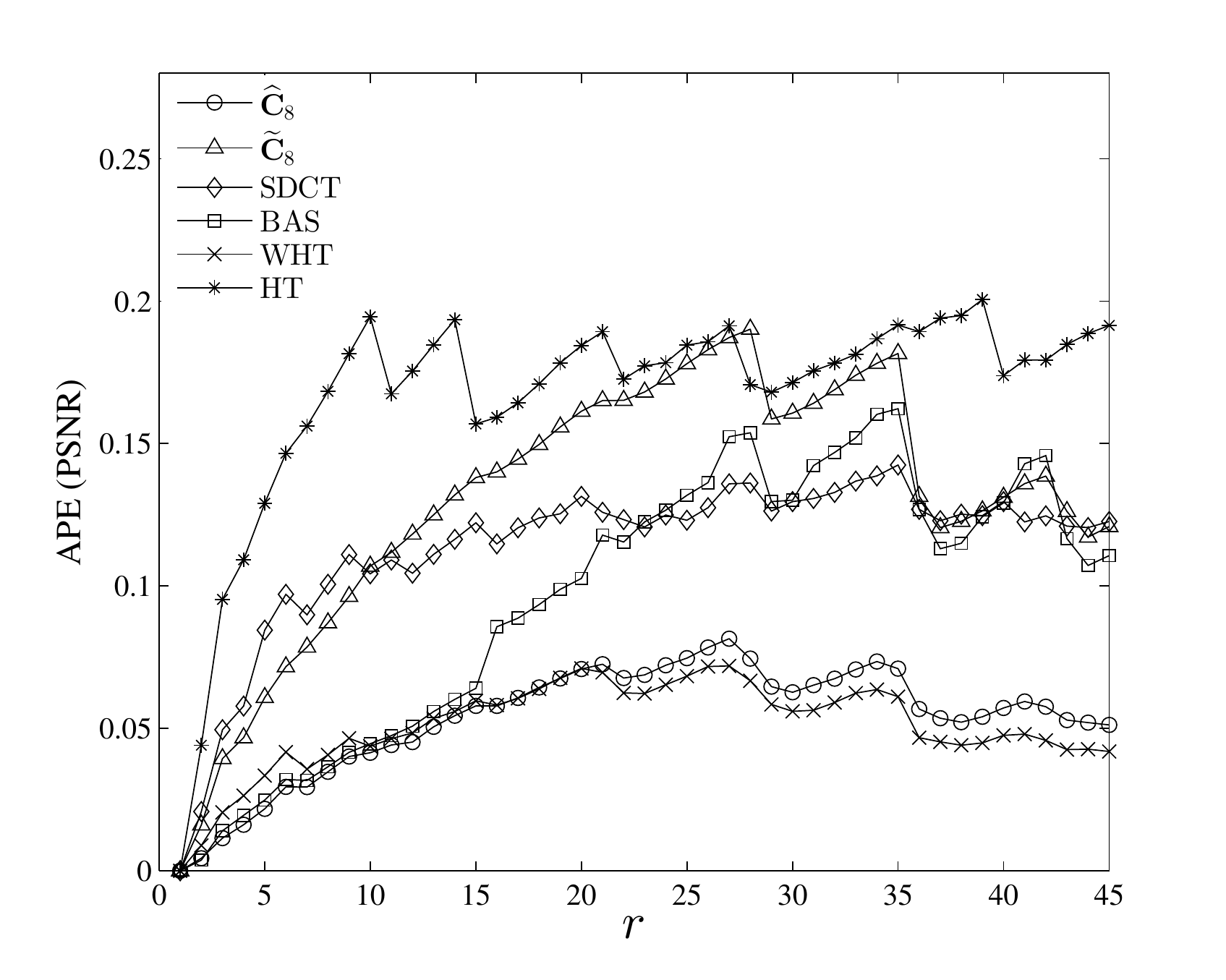}\label{fig:subfig3.2}}
\caption[]
{PSNR measures for several compression ratios.}
\label{figura3}
\centering
\subfigure[Average SSIM]
{\includegraphics[width=0.49\linewidth]{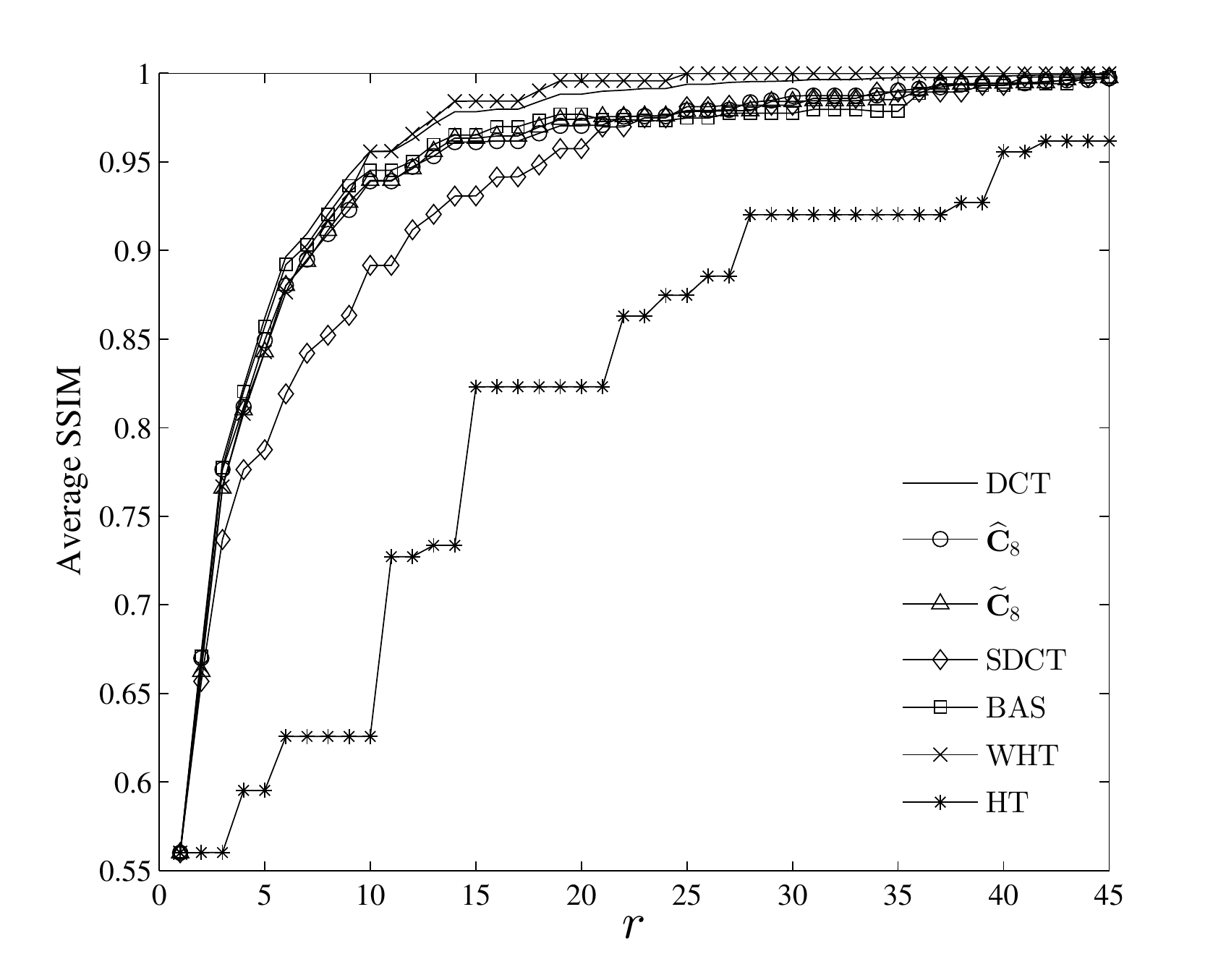}\label{fig:subfig4.1}}
\subfigure[Average SSIM absolute percentage error relative to the DCT]
{\includegraphics[width=0.49\linewidth]{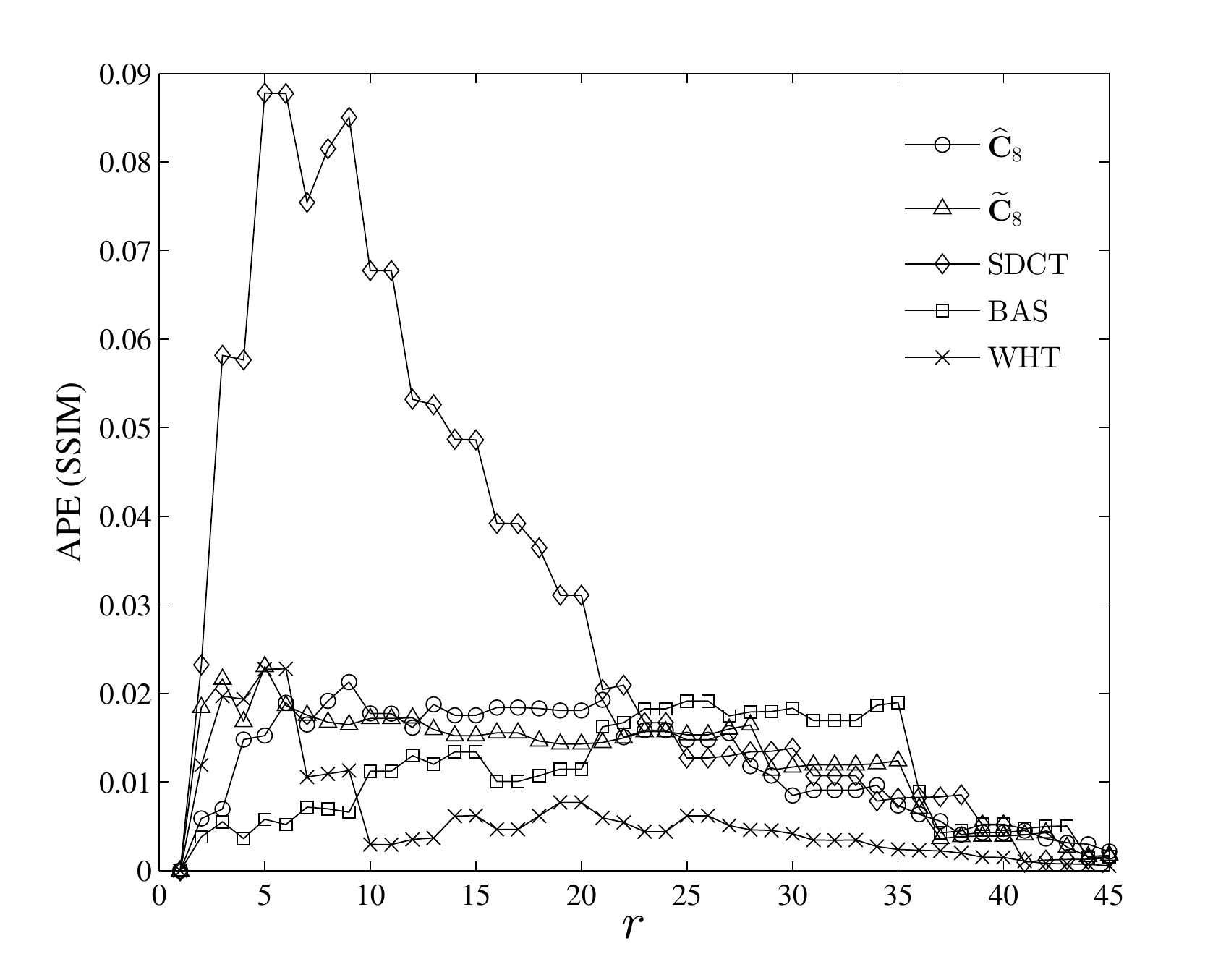}\label{fig:subfig4.2}}
\caption[]
{SSIM measures for several compression ratios.}
\label{figura4}
\end{figure}

Results
of the still image compression experiment
are presented
in
Figure~\ref{figura3} and \ref{figura4}.
In Figure~\ref{fig:subfig4.2},
the curve corresponding
to the HT
was suppressed
because
it presents
excessively high values in comparison
with other
curves.
In terms of PSNR,
$\widehat{\mathbf{C}}_8$
outperforms both the SDCT and the BAS approximations;
and provides similar results as to those furnished by
the
WHT,
but at a lower computational cost.
For SSIM results,
both transforms $\widetilde{\mathbf{C}}_8$ and $\widehat{\mathbf{C}}_8$
show similar results.
In terms of PSNR,
the proposed approximation
$\widehat{\mathbf{C}}_8$
performs closely to the WHT.
Figure~\ref{fig:subfig3.1} and \ref{fig:subfig3.2}
show that
$\widehat{\mathbf{C}}_8$
is superior to the WHT
at high compression ratios ($r\leq 15$).

\begin{figure}[t]
\centering
\subfigure[DCT]{
\includegraphics[width=0.3\linewidth]{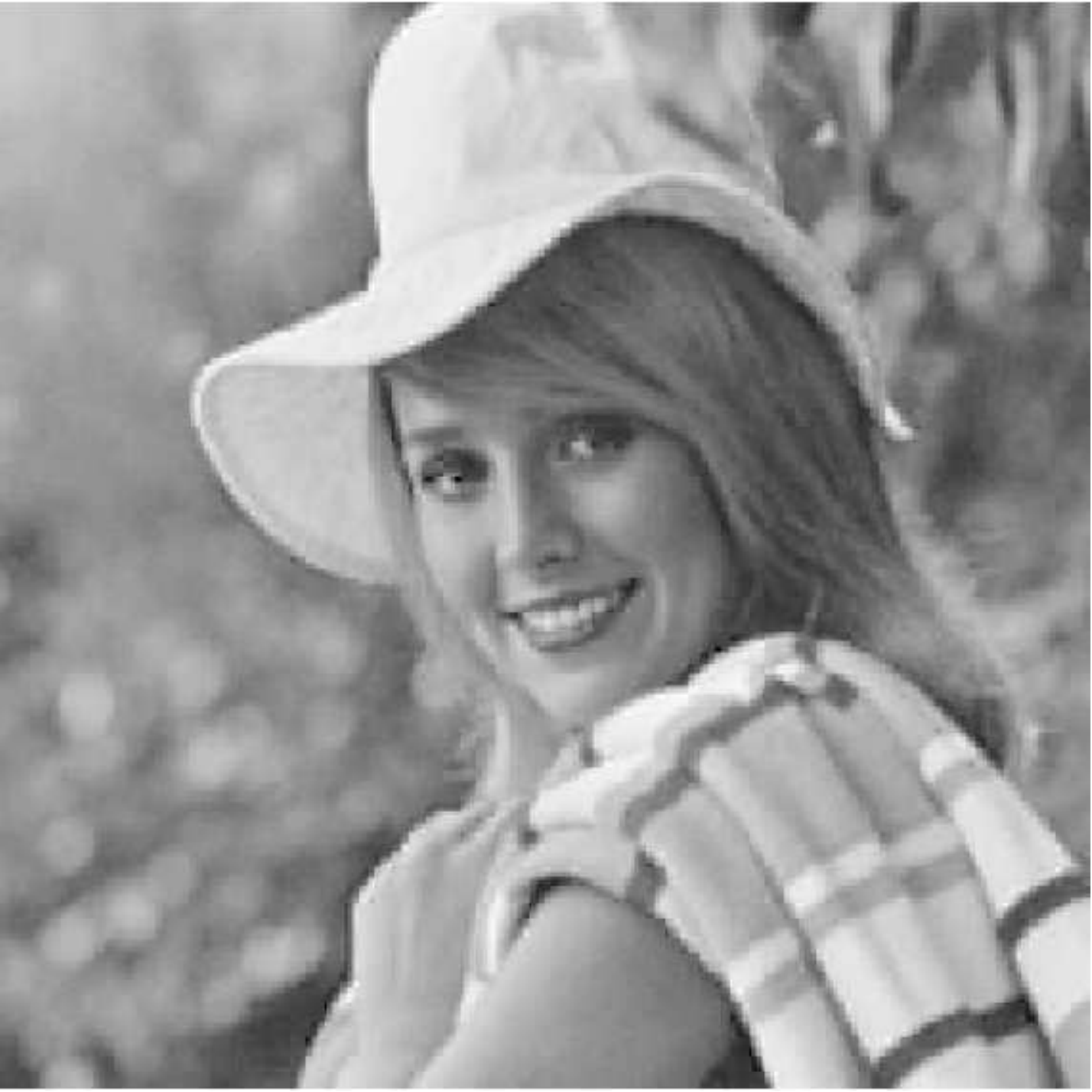}}
\label{fig:subfig5.1}
\subfigure[SDCT]{
\includegraphics[width=0.3\linewidth]{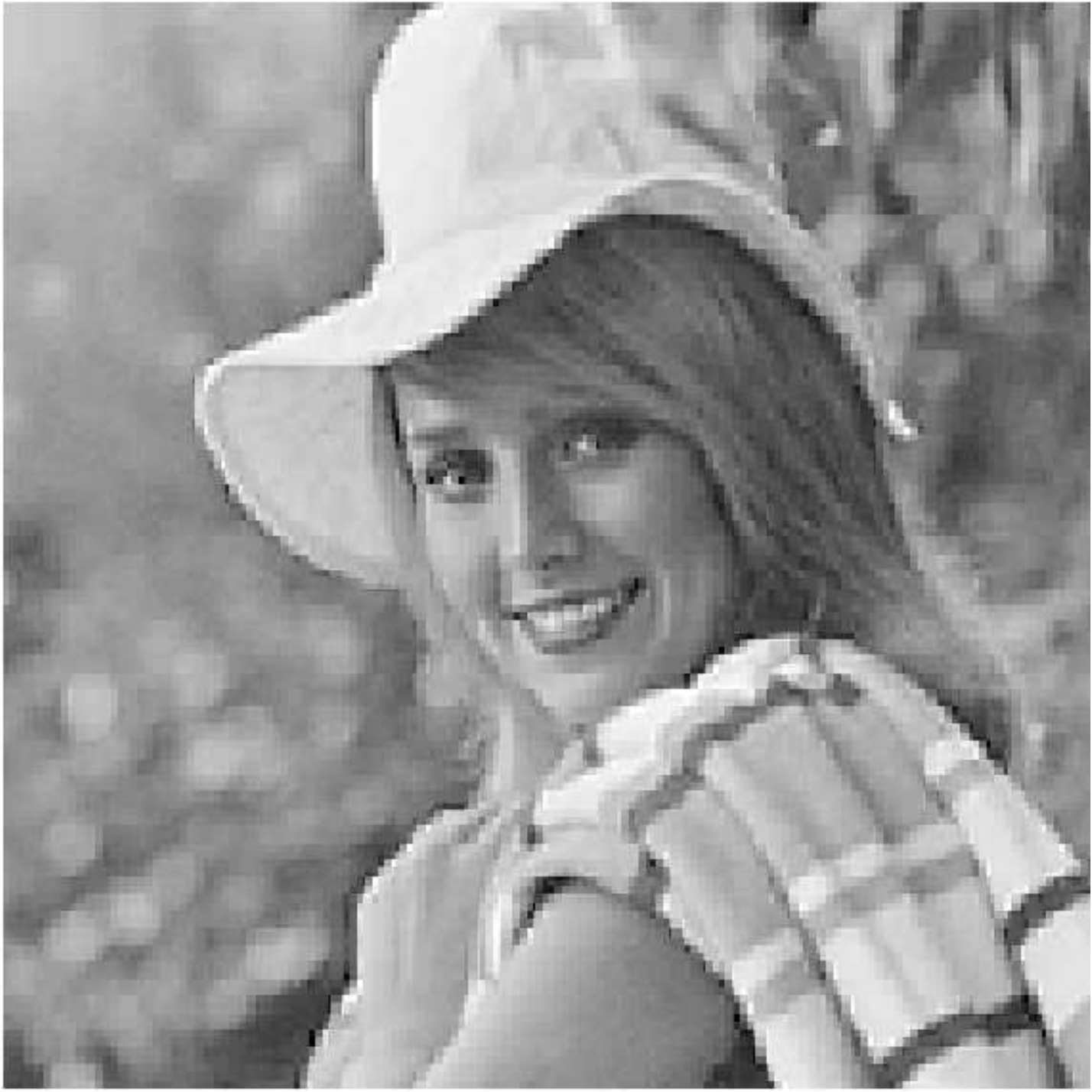}}
\label{fig:subfig5.2}
\subfigure[WHT]{
\includegraphics[width=0.3\linewidth]{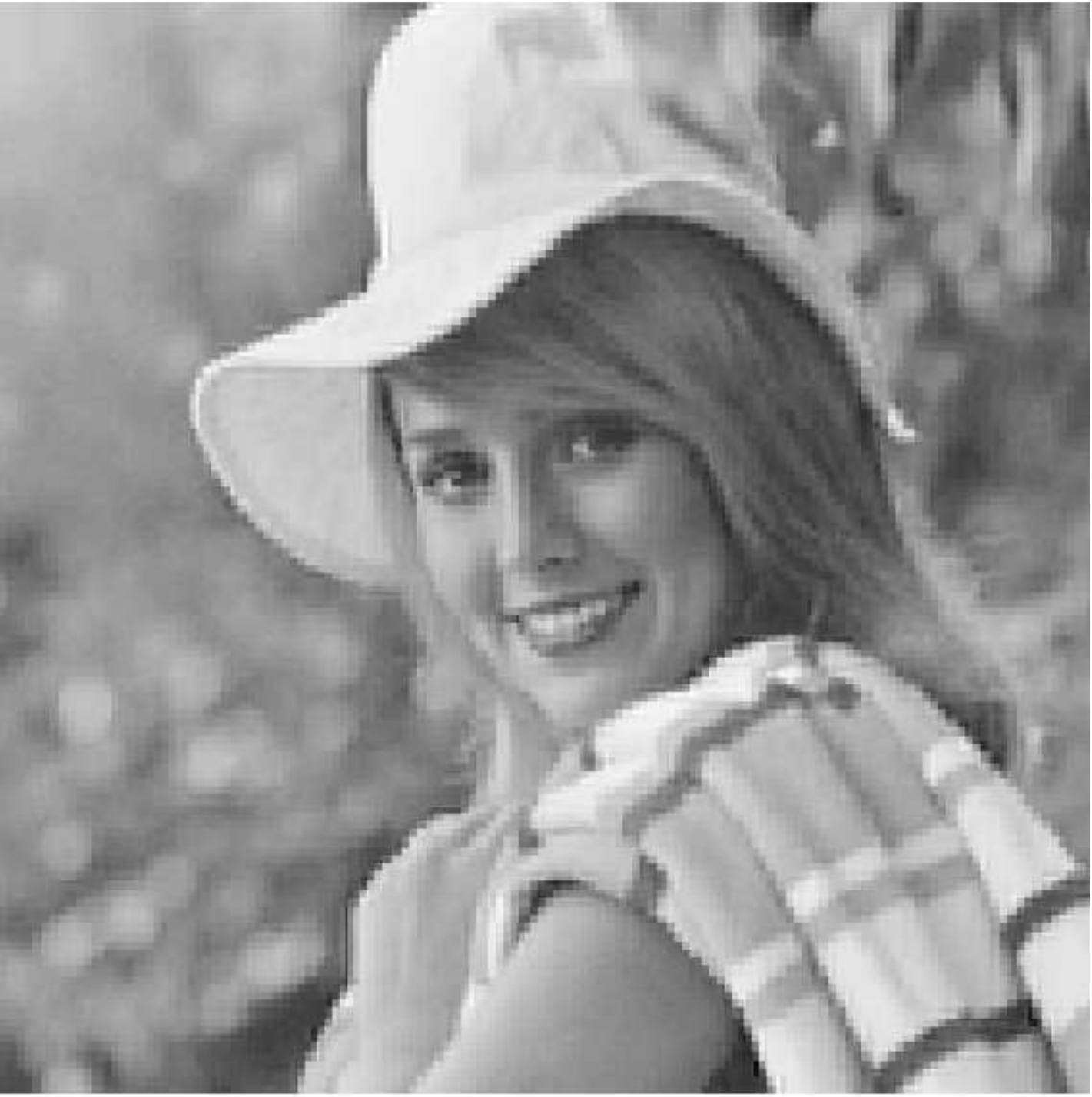}}
\label{fig:subfig5.3}
\subfigure[HT]{
\includegraphics[width=0.3\linewidth]{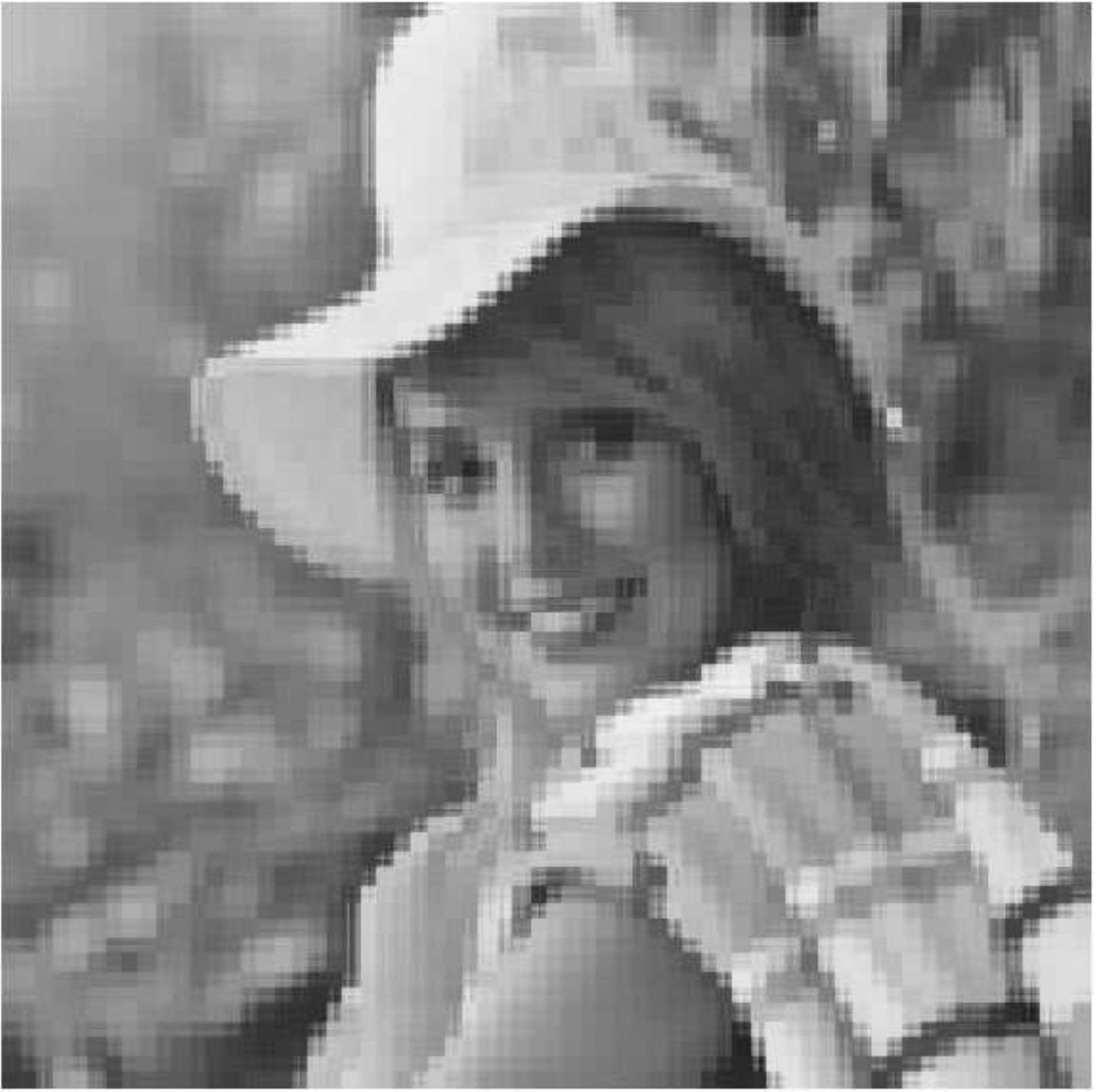}}
\label{fig:subfig5.4}
\subfigure[BAS]{
\includegraphics[width=0.3\linewidth]{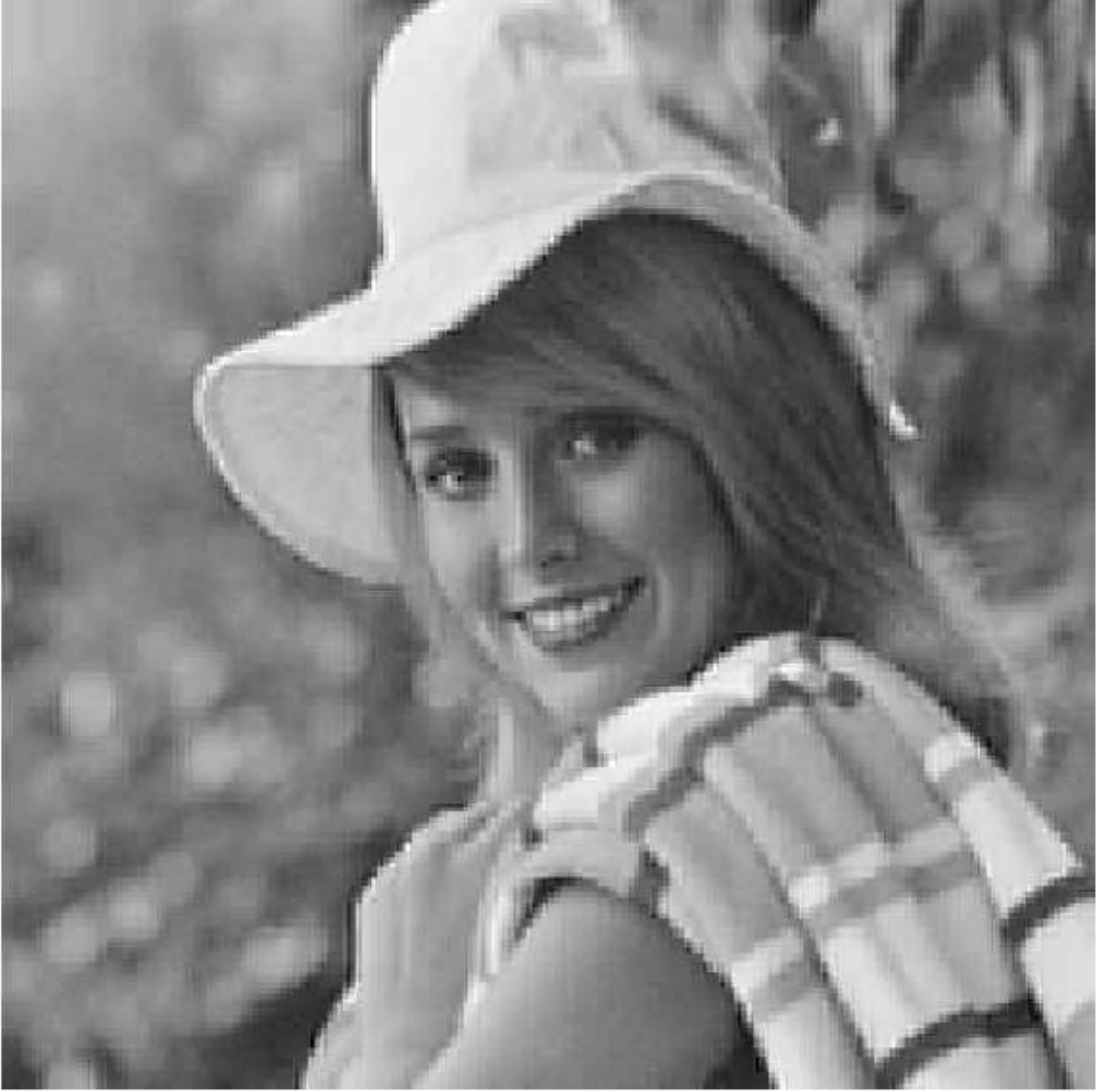}}
\label{fig:subfig5.5}
\subfigure[$\widehat{\mathbf{C}}_8$]{
\includegraphics[width=0.3\linewidth]{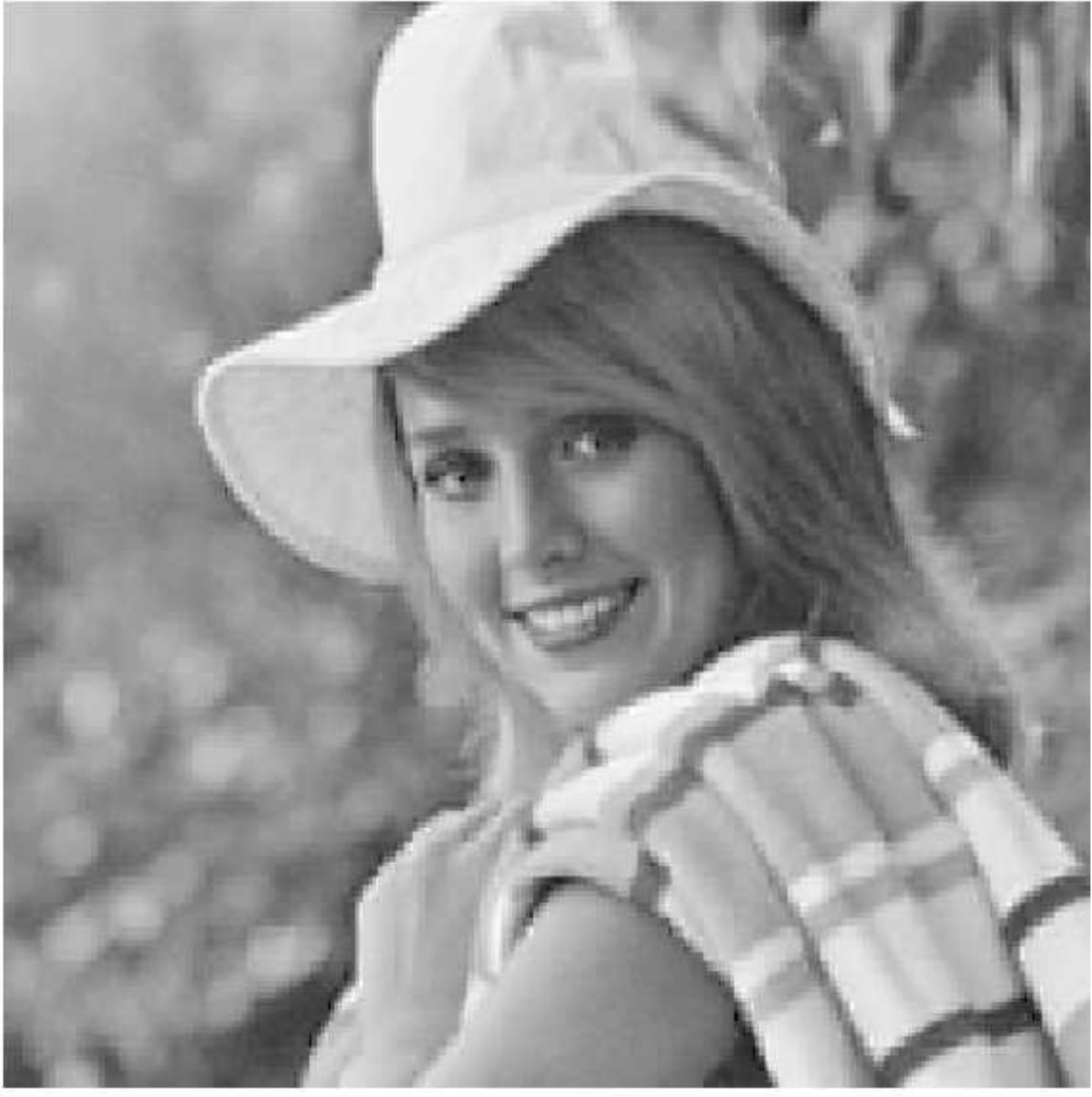}}
\label{fig:subfig5.6}
\caption[]{\label{figura5}\texttt{Elaine} image
 ($r=6$):
\subref{fig:subfig5.1}~DCT,
\subref{fig:subfig5.2}~SDCT,
\subref{fig:subfig5.3}~WHT,
\subref{fig:subfig5.4}~HT,
\subref{fig:subfig5.5}~BAS
and
\subref{fig:subfig5.6}~$\widehat{\mathbf{C}}_8$.}
\end{figure}

A qualitative analysis
is displayed
in
Figure~\ref{figura5}.
Standard \texttt{Elaine} image
was compressed
and reconstructed
according to
the
SDCT, WHT, HT, BAS,
and
the
proposed approximation $\widehat{\mathbf{C}}_8$
for visual inspection.
Compressed image resulted from
the
exact DCT
is also
exhibited.
All images were compressed with $r=6$,
which represents
a removal
of approximately $90.6\%$ of
the transformed
coefficients.
The visual analysis of the obtained images
shows the superiority of the
proposed transform~$\widehat{\mathbf{C}}_8$
over the SDCT in image compression.
Furthermore,
Table~\ref{tabela3}
lists the PSNR and SSIM values
for the \texttt{Elaine} image,
for $r=6$.
The PSNR and SSIM values are listed for two additional images (\texttt{Lenna} and \texttt{Boat} images).
The proposed
transform
$\widehat{\mathbf{C}}_8$
outperforms the HT, WHT, and SDCT approximations in terms of PSNR and SSIM and
BAS approximation in terms of PSNR for the
\texttt{Elaine} and \texttt{Lenna} images.

\begin{table}
\caption{Quality measures for three compressed images, considering $r=6$}
\label{tabela3}
\begin{center}
\begin{tabular}{lcccccc}
\toprule
&\multicolumn{2}{c}{\texttt{Elaine} image}&\multicolumn{2}{c}{\texttt{Lenna} image}&\multicolumn{2}{c}{\texttt{Boat} image}\\
\midrule
Transform & PSNR & SSIM & PSNR & SSIM & PSNR & SSIM\\
\midrule
DCT~\cite{Chen1977} &$31.03$&$0.95$&$29.90$&$0.95$&$26.94$&$0.92$\\
$\widehat{\mathbf{C}}_8$&$30.00$&$0.94$&$28.79$&$0.94$&$26.04$&$0.91$\\
BAS~\cite{Bouguezel2008} &$29.45$&$0.94$&$28.28$&$0.95$&$26.13$&$0.92$\\
WHT~\cite{Horadam2010} &$28.91$&$0.92$&$27.93$&$0.92$&$25.85$&$0.90$\\
SDCT~\cite{Haweel2001} &$27.59$&$0.88$&$26.26$&$0.88$&$24.09$&$0.85$\\
HT~\cite{Seberry2005} &$25.44$&$0.77$&$24.27$&$0.75$&$24.27$&$0.68$\\
\bottomrule
\end{tabular}
\end{center}
\end{table}

\subsection{Blocking artifact analysis}

\begin{figure}[t]
\centering
\subfigure[WHT]{
\includegraphics[width=0.31\linewidth]{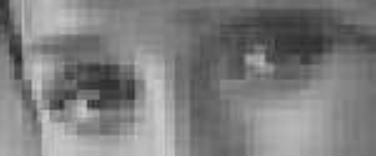}}
\label{fig:subfig6.2}
\subfigure[BAS]{
\includegraphics[width=0.305\linewidth]{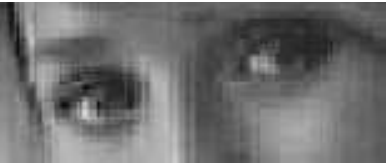}}
\label{fig:subfig6.3}
\subfigure[$\widehat{\mathbf{C}}_8$]{
\includegraphics[width=0.315\linewidth]{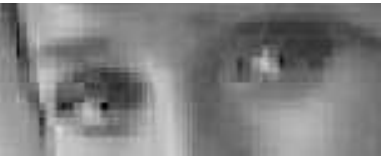}}
\label{fig:subfig6.1}
\caption[]{\label{figura6}\emph{Blocking artifact} effect in the \texttt{Elaine} image ($r=6$):
\subref{fig:subfig6.2}~WHT,
\subref{fig:subfig6.3}~BAS
and
\subref{fig:subfig6.1}~$\widehat{\mathbf{C}}_8$. }
\end{figure}

A visually undesirable effect in image compression
is the emergence of blocking artifacts~\cite[p.~573]{Gonzalez2006}.
Figure~\ref{figura6}
shows a qualitative
comparison
in terms of blocking artifact
resulted from $\widehat{\mathbf{C}}_8$, WHT, and BAS.
The proposed approximation $\widehat{\mathbf{C}}_8$
effected
a lower
degree of blocking artifact
comparatively with the WHT
and BAS.

\section{HEVC-compliant video encoding}
\label{s:video_coding}

In this section,
we aim at demonstrating
the practical real-world
applicability
of the proposed DCT approximations
for
video coding.
However,
the HEVC standard requires
not only an 8-point transform, but also 4-, 16-, and 32-point transforms.
In order to derive
Chen's DCT approximations
for larger blocklengths,
we considered
the
algorithm proposed by Jridi-Alfalou-Meher (JAM)~\cite{Jridi2015}.
We embedded
these low-complexity transforms
into a publicly available reference software
compliant with the HEVC standard~\cite{HEVCReference}.

The JAM algorithm consists of a scalable method for obtaining
higher order transforms from an 8-point DCT approximation.
An $N$-point DCT approximation~$\check{\mathbf{T}}_N$,
where $N$ is a power of two,
is recursively obtained through:
\begin{align*}
 \check{\mathbf{T}}_N = \frac{1}{\sqrt{2}} \mathbf{M}^\text{per}_N \begin{bmatrix}
\check{\mathbf{T}}_{\frac{N}{2}} &  \mathbf{0}_{\frac{N}{2}}\\
\mathbf{0}_{\frac{N}{2}} & \check{\mathbf{T}}_{\frac{N}{2}}
\end{bmatrix} \mathbf{M}^\text{add}_N,
\end{align*}
where
\begin{align*}
\mathbf{M}^\text{add}_N =  \begin{bmatrix}
\mathbf{I}_{\frac{N}{2}} &  \mathbf{\bar{I}}_{\frac{N}{2}}\\
\mathbf{\bar{I}}_{\frac{N}{2}} & -\mathbf{I}_{\frac{N}{2}}
\end{bmatrix}
\quad
\text{ and }
\quad
 \mathbf{M}^\text{per}_N =  \begin{bmatrix}
\mathbf{P}_{N-1,\frac{N}{2}} &  \mathbf{0}_{1,\frac{N}{2}}\\
\mathbf{0}_{1,\frac{N}{2}} & \mathbf{P}_{N-1,\frac{N}{2}}
\end{bmatrix},
\end{align*}
and
$\mathbf{P}_{N-1,\frac{N}{2}}$
is an $(N-1) \times (N/2)$
resulting
from expanding
the
identity matrix~$\mathbf{I}_\frac{N}{2}$
by interlacing it
with
zero rows.
The matrix $\mathbf{M}^\text{per}_N$ is
a permutation matrix and does not introduce
any arithmetic cost.
Matrix $\mathbf{M}^\text{add}_N$ contributes with only $N$ additions.
Furthermore,
the scaling factor $1/\sqrt{2}$
can be merged into the image compression
quantization step and
does not contribute to the arithmetic complexity of the transform.
Thus,
the additive complexity of $\mathbf{\check{T}}_N$
is equal to
twice
the additive complexity of
$\mathbf{\check{T}}_{\frac{N}{2}}$
plus
$N$~additions~\cite{Jridi2015}.

\subsection{Chen's DCT approximations for large blocklengths}

In its original form,
the JAM algorithm
employs the 8-point DCT approximation
introduced in~\cite{Cintra2011}.
In this section,
we adapt the JAM algorithm
to derive
DCT approximations
based on Chen's algorithm
for
arbitrary
power-of-two ($N>8$) blocklengths.
We are specially interested in 16- and 32-point
low-complexity transformations
for subsequent embedding into the HEVC standard.
Let $N>8$ be a power of two.
We introduce
the Chen's signed and rounded transformations, respectively,
according to the following recursion:
\begin{align}
\label{ec:app_N}
 \widetilde{\mathbf{T}}_N = \mathbf{M}^\text{per}_N \begin{bmatrix}
\widetilde{\mathbf{T}}_{\frac{N}{2}} &  \mathbf{0}_{\frac{N}{2}}\\
\mathbf{0}_{\frac{N}{2}} & \widetilde{\mathbf{T}}_{\frac{N}{2}}
\end{bmatrix} \mathbf{M}^\text{add}_N
\quad
 \text{ and }
\quad
 \widehat{\mathbf{T}}_N = \mathbf{M}^\text{per}_N \begin{bmatrix}
\widehat{\mathbf{T}}_{\frac{N}{2}} &  \mathbf{0}_{\frac{N}{2}}\\
\mathbf{0}_{\frac{N}{2}} & \widehat{\mathbf{T}}_{\frac{N}{2}}
\end{bmatrix} \mathbf{M}^\text{add}_N.
\end{align}
Based on~\eqref{ec:ecuation4} and~\eqref{ec:ecuation5},
$\widetilde{\mathbf{T}}_{\frac{N}{2}}$ and $\widehat{\mathbf{T}}_{\frac{N}{2}}$ admit
the following factorizations:
\begin{align*}
 \widetilde{\mathbf{T}}_{\frac{N}{2}} &= \check{\mathbf{P}}_{\frac{N}{2}}\,\mathbf{M}_{\frac{N}{2}}^{(1)}\,\,\widetilde{\mathbf{M}}_{\frac{N}{2}}^{(2)}\,\,\widetilde{\mathbf{M}}_{\frac{N}{2}}^{(3)}\,\,\widetilde{\mathbf{M}}_{\frac{N}{2}}^{(4)}\,\, \check{\mathbf{B}}_{\frac{N}{2}},\\
 \widehat{\mathbf{T}}_{\frac{N}{2}} &= \check{\mathbf{P}}_{\frac{N}{2}}\,\mathbf{M}_{\frac{N}{2}}^{(1)}\,\,\widehat{\mathbf{M}}_{\frac{N}{2}}^{(2)}\,\,\widehat{\mathbf{M}}_{\frac{N}{2}}^{(3)}\,\,\widehat{\mathbf{M}}_{\frac{N}{2}}^{(4)}\,\, \check{\mathbf{B}}_{\frac{N}{2}}.
\end{align*}
Thus, applying the factorizations above in~\eqref{ec:app_N} and expanding them, we obtain:
\begin{align}
\label{eq:transformation_N_sign}
 \widetilde{\mathbf{T}}_{N} &= \check{\mathbf{P}}_{N}\,\mathbf{M}_{N}^{(1)}\,\,\widetilde{\mathbf{M}}_{N}^{(2)}\,\,\widetilde{\mathbf{M}}_{N}^{(3)}\,\,\widetilde{\mathbf{M}}_{N}^{(4)}\,\, \check{\mathbf{B}}_{N},\\
 \label{eq:transformation_N_round}
 \widehat{\mathbf{T}}_{N} &= \check{\mathbf{P}}_{N}\,\mathbf{M}_{N}^{(1)}\,\,\widehat{\mathbf{M}}_{N}^{(2)}\,\,\widehat{\mathbf{M}}_{N}^{(3)}\,\,\widehat{\mathbf{M}}_{N}^{(4)}\,\, \check{\mathbf{B}}_{N},
\end{align}
where
\begin{align*}
 \check{\mathbf{P}}_{N} &= \mathbf{M}^\text{per}_N \begin{bmatrix}
\check{\mathbf{P}}_{\frac{N}{2}} &  \mathbf{0}_{\frac{N}{2}}\\
\mathbf{0}_{\frac{N}{2}} & \check{\mathbf{P}}_{\frac{N}{2}}
\end{bmatrix},\quad \check{\mathbf{B}}_{N} = \begin{bmatrix}
\check{\mathbf{B}}_{\frac{N}{2}} &  \mathbf{0}_{\frac{N}{2}}\\
\mathbf{0}_{\frac{N}{2}} & \check{\mathbf{B}}_{\frac{N}{2}}
\end{bmatrix} \mathbf{M}^\text{add}_N,\quad \mathbf{M}_{N}^{(1)} = \begin{bmatrix}
\mathbf{M}_{\frac{N}{2}}^{(1)} &  \mathbf{0}_{\frac{N}{2}}\\
\mathbf{0}_{\frac{N}{2}} & \mathbf{M}_{\frac{N}{2}}^{(1)}
\end{bmatrix},\\
\widetilde{\mathbf{M}}_{N}^{(i)} &= \begin{bmatrix}
\widetilde{\mathbf{M}}_{\frac{N}{2}}^{(i)} &  \mathbf{0}_{\frac{N}{2}}\\
\mathbf{0}_{\frac{N}{2}} & \widetilde{\mathbf{M}}_{\frac{N}{2}}^{(i)}
\end{bmatrix},\quad \widehat{\mathbf{M}}_{N}^{(i)} = \begin{bmatrix}
\widehat{\mathbf{M}}_{\frac{N}{2}}^{(i)} &  \mathbf{0}_{\frac{N}{2}}\\
\mathbf{0}_{\frac{N}{2}} & \widehat{\mathbf{M}}_{\frac{N}{2}}^{(i)}
\end{bmatrix},\quad i=2,3,4.
\end{align*}
Their inverse transformations possess the following factorizations:
\begin{align}
\label{eq:transformation_inverse_N_sign}
 \widetilde{\mathbf{T}}_{N}^{-1} &= \frac{4}{N}\,\check{\mathbf{B}}_{N}^\top\,\left.\widetilde{\mathbf{M}}_{N}^{(4)}\right.^{-1}\left.\widetilde{\mathbf{M}}_{N}^{(3)}\right.^{-1}\left.\widetilde{\mathbf{M}}_{N}^{(2)}\right.^{-1}\left.\mathbf{M}_{N}^{(1)}\right.^\top\check{\mathbf{P}}_{N}^\top,\\
 \label{eq:transformation_inverse_N_round}
 \widehat{\mathbf{T}}_{N}^{-1} &= \frac{4}{N}\,\check{\mathbf{B}}_{N}^\top\,\left.\widehat{\mathbf{M}}_{N}^{(4)}\right.^{-1}\left.\widehat{\mathbf{M}}_{N}^{(3)}\right.^{-1}\left.\widehat{\mathbf{M}}_{N}^{(2)}\right.^{-1}\left.\mathbf{M}_{N}^{(1)}\right.^\top\check{\mathbf{P}}_{N}^\top,
\end{align}
where
\begin{align*}
 \check{\mathbf{P}}_{N}^\top &= \begin{bmatrix}
\check{\mathbf{P}}_{\frac{N}{2}}^\top &  \mathbf{0}_{\frac{N}{2}}\\
\mathbf{0}_{\frac{N}{2}} & \check{\mathbf{P}}_{\frac{N}{2}}^\top
\end{bmatrix}\left.\mathbf{M}^\text{per}_N\right.^\top,\quad \check{\mathbf{B}}_{N}^\top = \left.\mathbf{M}^\text{add}_N\right.^\top\begin{bmatrix}
\check{\mathbf{B}}_{\frac{N}{2}}^\top &  \mathbf{0}_{\frac{N}{2}}\\
\mathbf{0}_{\frac{N}{2}} & \check{\mathbf{B}}_{\frac{N}{2}}^\top
\end{bmatrix},\quad \left.\mathbf{M}_{N}^{(1)}\right.^\top = \begin{bmatrix}
\left.\mathbf{M}_{\frac{N}{2}}^{(1)}\right.^\top &  \mathbf{0}_{\frac{N}{2}}\\
\mathbf{0}_{\frac{N}{2}} & \left.\mathbf{M}_{\frac{N}{2}}^{(1)}\right.^\top
\end{bmatrix},\\
\left.\widetilde{\mathbf{M}}_{N}^{(i)}\right.^{-1} &= \begin{bmatrix}
\left.\widetilde{\mathbf{M}}_{\frac{N}{2}}^{(i)}\right.^{-1} &  \mathbf{0}_{\frac{N}{2}}\\
\mathbf{0}_{\frac{N}{2}} & \left.\widetilde{\mathbf{M}}_{\frac{N}{2}}^{(i)}\right.^{-1}
\end{bmatrix},\quad \left.\widehat{\mathbf{M}}_{N}^{(i)}\right.^{-1} = \begin{bmatrix}
\left.\widehat{\mathbf{M}}_{\frac{N}{2}}^{(i)}\right.^{-1} &  \mathbf{0}_{\frac{N}{2}}\\
\mathbf{0}_{\frac{N}{2}} & \left.\widehat{\mathbf{M}}_{\frac{N}{2}}^{(i)}\right.^{-1}
\end{bmatrix},\quad i=2,3,4.
\end{align*}
In particular,
for $N=16$,
we have from~\eqref{ec:ecuation4} and~\eqref{ec:ecuation5}
that
$\check{\mathbf{P}}_8=\mathbf{P}_8$, $\check{\mathbf{B}}_8=\mathbf{B}_8$, $\mathbf{M}_8^{(1)}=\mathbf{M}_1$,
$\widetilde{\mathbf{M}}_8^{(i)}=\widetilde{\mathbf{M}}_{i}$, $\widehat{\mathbf{M}}_8^{(i)}=\widehat{\mathbf{M}}_{i}$,
$i=2,3,4$,
and therefore
it yields:
\begin{align*}
 \widetilde{\mathbf{T}}_{16} &= \check{\mathbf{P}}_{16}\,\mathbf{M}_{16}^{(1)}\,\,\widetilde{\mathbf{M}}_{16}^{(2)}\,\,\widetilde{\mathbf{M}}_{16}^{(3)}\,\,\widetilde{\mathbf{M}}_{16}^{(4)}\,\, \check{\mathbf{B}}_{16},\\
 \widehat{\mathbf{T}}_{16} &= \check{\mathbf{P}}_{16}\,\mathbf{M}_{16}^{(1)}\,\,\widehat{\mathbf{M}}_{16}^{(2)}\,\,\widehat{\mathbf{M}}_{16}^{(3)}\,\,\widehat{\mathbf{M}}_{16}^{(4)}\,\, \check{\mathbf{B}}_{16},
\end{align*}
where
\begin{align*}
 \check{\mathbf{P}}_{16} &= \mathbf{M}^\text{per}_{16} \begin{bmatrix}
\mathbf{P}_8 &  \mathbf{0}_8\\
\mathbf{0}_8 & \mathbf{P}_8
\end{bmatrix},\quad \check{\mathbf{B}}_{16} = \begin{bmatrix}
\mathbf{B}_8 &  \mathbf{0}_8\\
\mathbf{0}_8 & \mathbf{B}_8
\end{bmatrix} \mathbf{M}^\text{add}_{16},\quad \mathbf{M}_{16}^{(1)} = \begin{bmatrix}
\mathbf{M}_1 &  \mathbf{0}_8\\
\mathbf{0}_8 & \mathbf{M}_1
\end{bmatrix},\\
\widetilde{\mathbf{M}}_{16}^{(i)} &= \begin{bmatrix}
\widetilde{\mathbf{M}}_{i} &  \mathbf{0}_8\\
\mathbf{0}_8 & \widetilde{\mathbf{M}}_{i}
\end{bmatrix},\quad \widehat{\mathbf{M}}_{16}^{(i)} = \begin{bmatrix}
\widehat{\mathbf{M}}_{i} &  \mathbf{0}_8\\
\mathbf{0}_8 & \widehat{\mathbf{M}}_{i}
\end{bmatrix},\quad i=2,3,4.
\end{align*}

The near orthogonal DCT approximations linked to the proposed low-complexity matrices are given by
(cf. Section~\ref{sec: orthogonality}):
\begin{align*}
 \widetilde{\mathbf{C}}_N=\widetilde{\mathbf{S}}_N\,\widetilde{\mathbf{T}}_N,\quad \widehat{\mathbf{C}}_N=\widehat{\mathbf{S}}_N\,\widehat{\mathbf{T}}_N,
\end{align*}
where $\widetilde{\mathbf{S}}_N=\sqrt{[\operatorname{diag}(\widetilde{\mathbf{T}}_N\,\widetilde{\mathbf{T}}_N^\top)]^{-1}}$ and $\widehat{\mathbf{S}}_N=\sqrt{[\operatorname{diag}(\widehat{\mathbf{T}}_N\,\widehat{\mathbf{T}}_N^\top)]^{-1}}$.

Because the entries of
$\widetilde{\mathbf{T}}_8$ and $\widehat{\mathbf{T}}_8$
and their inverses
are in the set $\mathcal{P}=\{0,\pm\frac{1}{2},\pm1,\pm2\}$,
we have that the matrices in~\eqref{eq:transformation_N_sign},
\eqref{eq:transformation_N_round}, \eqref{eq:transformation_inverse_N_sign} and~\eqref{eq:transformation_inverse_N_round}
also have entries in~$\mathcal{P}$.
Moreover,
\eqref{eq:transformation_N_sign},
\eqref{eq:transformation_N_round}, \eqref{eq:transformation_inverse_N_sign} and~\eqref{eq:transformation_inverse_N_round}
can also be recursively obtained from~\eqref{ec:ecuation4}, \eqref{ec:ecuation5}, \eqref{ec:ecuation6}, and~\eqref{ec:ecuation7}.
Thus,
$\widetilde{\mathbf{C}}_N$ and $\widehat{\mathbf{C}}_N$ are low-complexity DCT approximations
for blocklength $N$.
The arithmetic complexity of the
proposed 16- and 32-point Chen's approximations
and
transformations
prescribed in the HEVC standard
are presented in Table~\ref{t:computational_cost-large}.
In terms of hardware implementation,
the circuitry corresponding to
$\widetilde{\mathbf{C}}_8$ and $\widehat{\mathbf{C}}_8$ and their inverses
can be re-used
for the hardware implementation of
both the direct and inverse Chen's DCT approximations
for larger blocklengths.

\begin{table}[t]%
\caption{Arithmetic complexity of the considered 16- and 32-point transforms}
\label{t:computational_cost-large}
\begin{center}
\begin{tabular}{lcccc}
\toprule
Transform & Mult & Add & Shift &Total \\
\midrule%
16-point exact DCT~\cite{Chen1977} & $44$ & $74$ & $0$ & $118$\\
16-point transform in HEVC \cite{Meher2014} & $0$ & $186$ & $86$ & $272$ \\
$\widehat{\mathbf{C}}_{16}$ & $0$ & $60$ & $0$ & $60$\\
$\widetilde{\mathbf{C}}_{16}$ & $0$ & $68$ & $0$ & $68$ \\
32-point exact DCT~\cite{Chen1977} & $116$ & $194$ & $0$ & $310$\\
32-point transform in HEVC \cite{Meher2014} & $0$ & $682$ & $278$ & $960$ \\
$\widehat{\mathbf{C}}_{32}$ & $0$ & $152$ & $0$ & $152$\\
$\widetilde{\mathbf{C}}_{32}$ & $0$ & $168$ & $0$ & $168$ \\
\bottomrule
\end{tabular}
\end{center}
\end{table}

\subsection{Results}

\begin{figure}
\centering
 \subfigure[]{\includegraphics[width=0.4\linewidth]{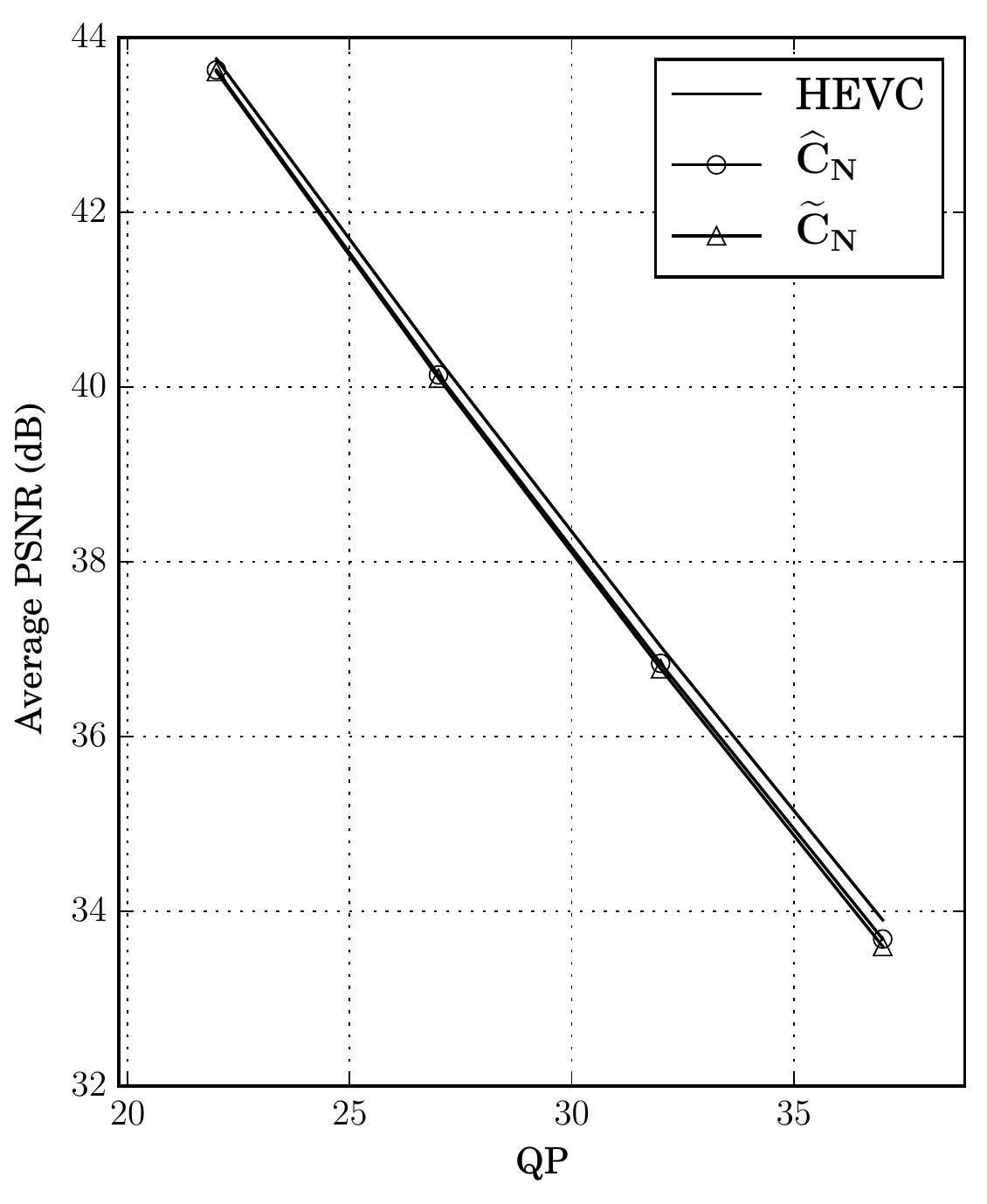}}
 \subfigure[]{\includegraphics[width=0.4\linewidth]{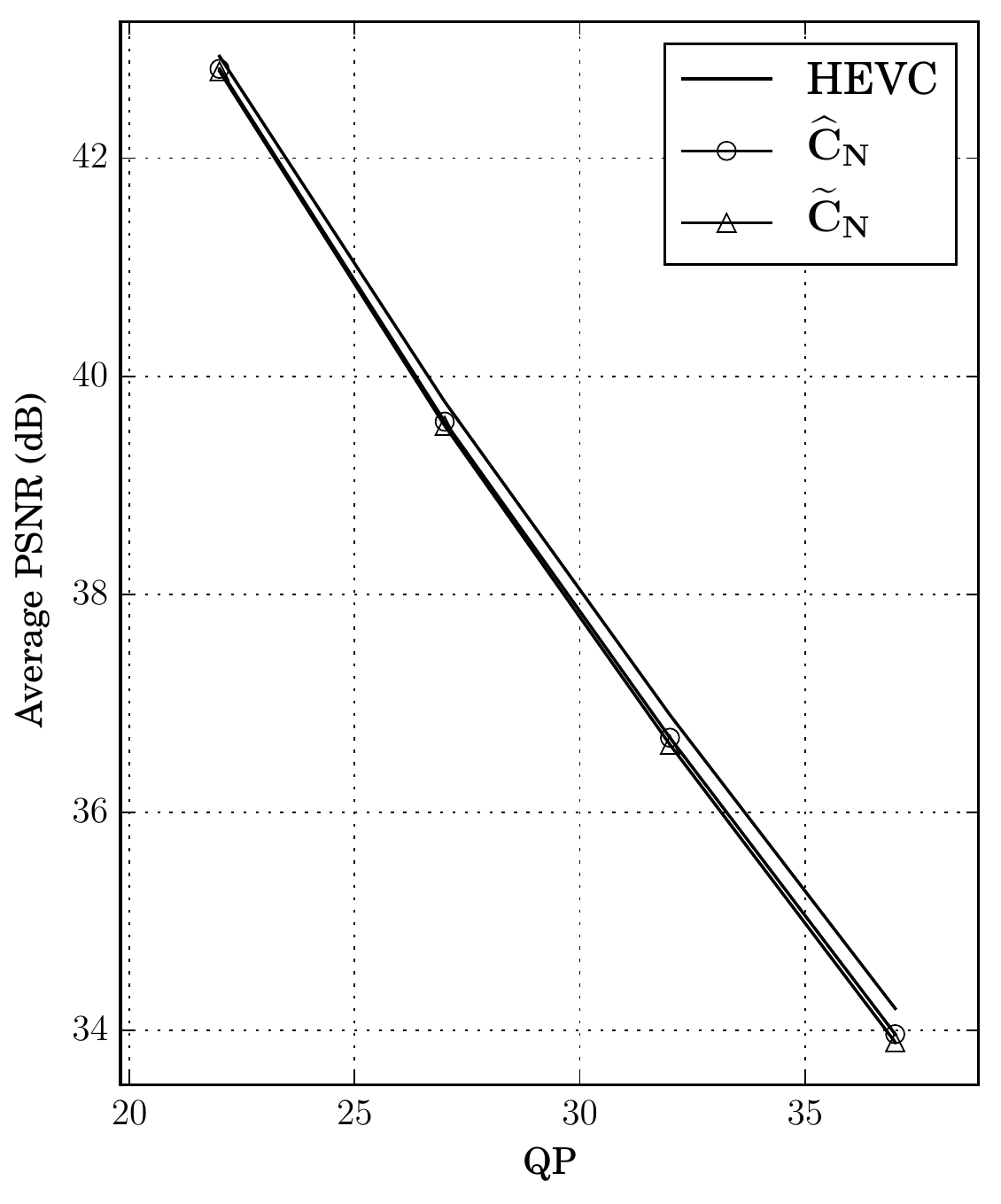}}
 \caption{
 Average PSNR for QP in \{22,27,32,37\} for videos in Groups (a)~A and (b)~B.}
\label{fig:figura7}
\end{figure}

The proposed approximations
were embedded into the HEVC reference software~\cite{HEVCReference}.
For video coding experiments
we considered two set of videos, namely:
(i)~Group A, which consider eleven CIF videos from~\cite{Xiph};
(ii)~Group B, with six standard video sequence,
one for each
class specified in the Common Test Conditions (CTC) document for HEVC~\cite{CTConditions2013}.
Such classification is based on the resolution, frame rate and, as consequence, the main applications of these kind of media~\cite{Naccari2014}.
All test parameters were set according to the CTC document,
including the quantization parameter (QP) that assumes values in \{22, 27, 32, 37\}.
As suggested in~\cite{Jridi2015}, we selected the \texttt{Main} profile and \texttt{All-Intra} mode for our experiments.
The PSNR measurements---already furnished by the reference software---were
obtained for each video frame
and YUV~channel.
The overall PSNR was obtained from each frame according to~\cite{Ohm2012}.
We averaged the PSNR  values for the first 100~frames of all videos in each group.
Figure~\ref{fig:figura7} shows
the average PSNR
in terms of
the quantization parameter (QP)
for each set of 8-, 16-, and 32-point transforms:
$\widetilde{\mathbf{C}}_N$,
$\widehat{\mathbf{C}}_N$,
and
the original transforms in the HEVC standard.
Results in Figure~\ref{fig:figura7}
show
no significant degradation in terms of PSNR
regardless of the video group.
The proposed approximations
resulted in essentially the same frame quality
while
having a very low computational cost when compared to
those originally employed in HEVC.

Additionally,
we computed the Bj\o ntegaard delta PSNR (BD-PSNR)
and delta rate (BD-Rate)~\cite{Hanhart2014, Bjontegaard2001}
for compressed videos
considering all discussed 8- to 32-point transformations.
The first 11~rows of Table~\ref{tab:videos} present the results for the Group~A whilst the remaining ones
correspond to Group~B.
We report a negligible impact in video quality
associated to the results from
the modified HEVC with the approximate transforms.
Similar to the still images experiments,
$\widehat{\mathbf{C}}_N$ performed better than $\widetilde{\mathbf{C}}_N$
with a degradation of no more than 0.70dB and 0.58dB for Groups~A and~B, respectively.
These declines in PSNR represent an increase of 10.63\% and 7.02\% in bitrate, respectively.

\begin{table*}[]
\centering
\caption{
Bj\o ntegaard metrics for the approximate transforms and tested video sequences}
\label{tab:videos}
\begin{tabular}{lccccc}
\toprule
\multirow{2}{*}{Video information}&\multicolumn{2}{c}{BD-PSNR (dB)}&\multicolumn{2}{c}{BD-Rate (\%)}\\
&$\widetilde{\mathbf{C}}_N$&$\widehat{\mathbf{C}}_N$&$\widetilde{\mathbf{C}}_N$&$\widehat{\mathbf{C}}_N$\\
\midrule
\texttt{Akiyo}&0.4600&0.2990&$-7.0310$&$-4.6870$\\
\texttt{Bowing}&0.5301&0.4316&$-7.4519$&$-6.1509$\\
\texttt{Coastguard}&0.7596&0.7026&$-11.3634$&$-10.6298$\\
\texttt{Container}&0.4075&0.3750&$-6.3002$&$-5.8044$\\
\texttt{Foreman}&0.2263&0.1627&$-4.4006$&$-3.2148$\\
\texttt{Hall\_monitor}&0.2754&0.1952&$-4.7577$&$-3.4125$\\
\texttt{Mobile}&0.2752&0.2629&$-2.7072$&$-2.5860$\\
\texttt{Mother\_daughter}&0.4202&0.3384&$-7.8362$&$-6.4112$\\
\texttt{News}&0.2539&0.1975&$-3.4211$&$-2.6772$\\
\texttt{Pamphlet}&0.4253&0.3680&$-5.8660$&$-5.1057$\\
\texttt{Silent}&0.3029&0.2399&$-5.7215$&$-4.6042$\\
\midrule
\texttt{PeopleOnStreet}&0.5350&0.4734&$-9.6530$&$-8.6227$\\
\texttt{BasketballDrive}&0.3372&0.2531&$-11.7780$&$-9.0093$\\
\texttt{RaceHorses}&0.6444&0.5781&$-7.7823$&$-7.0233$\\
\texttt{BlowingBubbles}&0.2563&0.1986&$-4.3438$&$-3.4080$\\
\texttt{KristenAndSara}&0.4651&0.3807&$-8.8416$&$-7.3234$\\
\texttt{BasketballDrillText}&0.1984&0.1565&$-3.7436$&$-2.9711$\\
\bottomrule
\end{tabular}
\end{table*}

\begin{figure}
\centering
\subfigure[HEVC]
{\includegraphics[width=.32\linewidth]{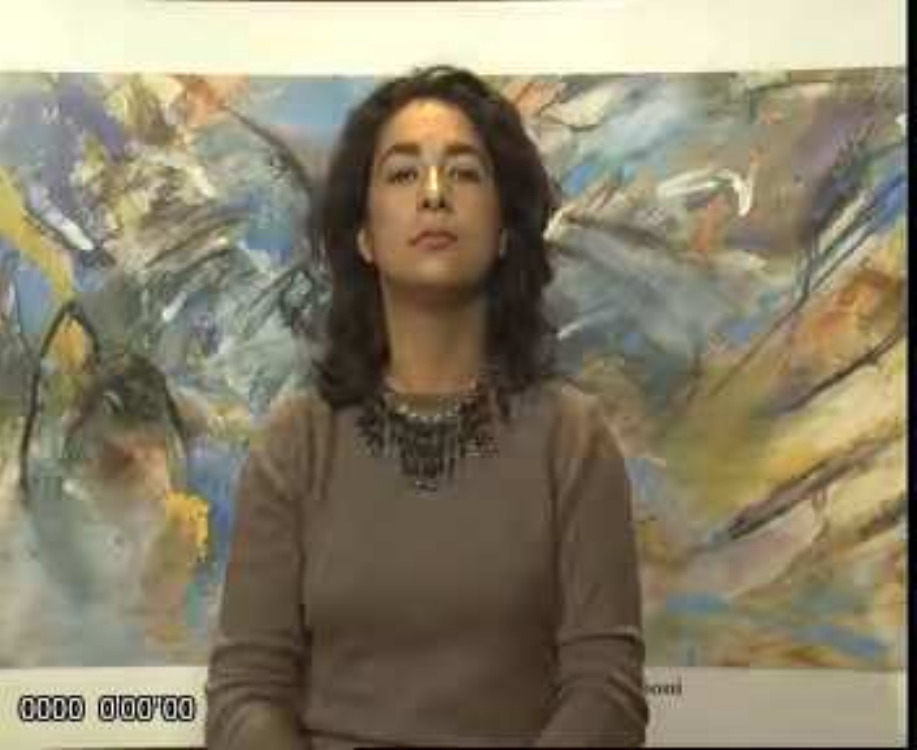}\label{fig:subfig8.1a}}
\subfigure[$\widetilde{\mathbf{C}}_N$]
{\includegraphics[width=.32\linewidth]{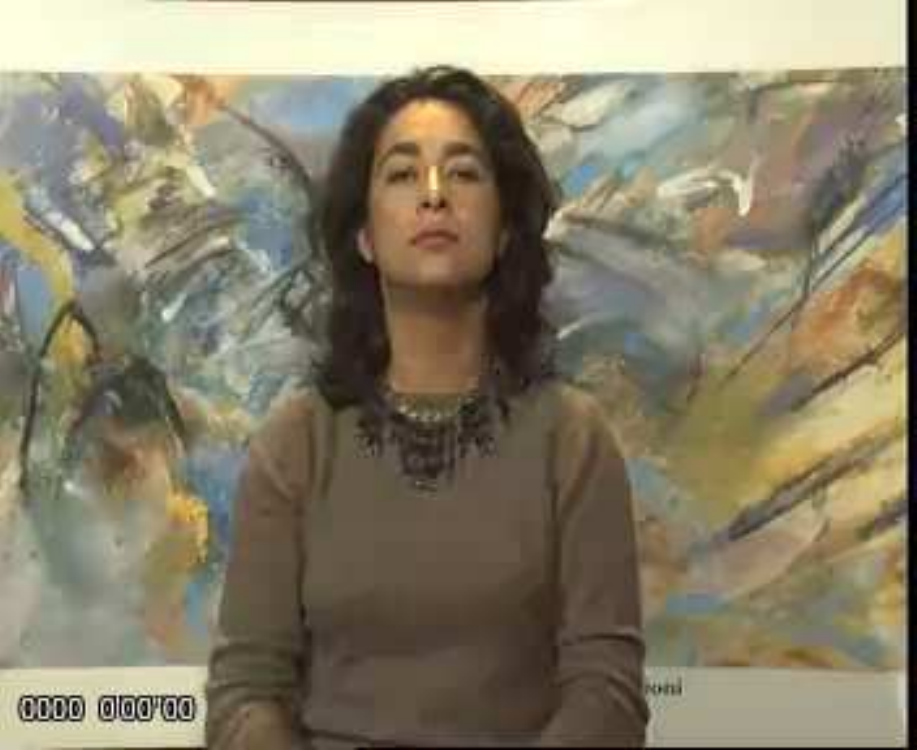}\label{fig:subfig8.2a}}
\subfigure[$\widehat{\mathbf{C}}_N$]
{\includegraphics[width=.32\linewidth]{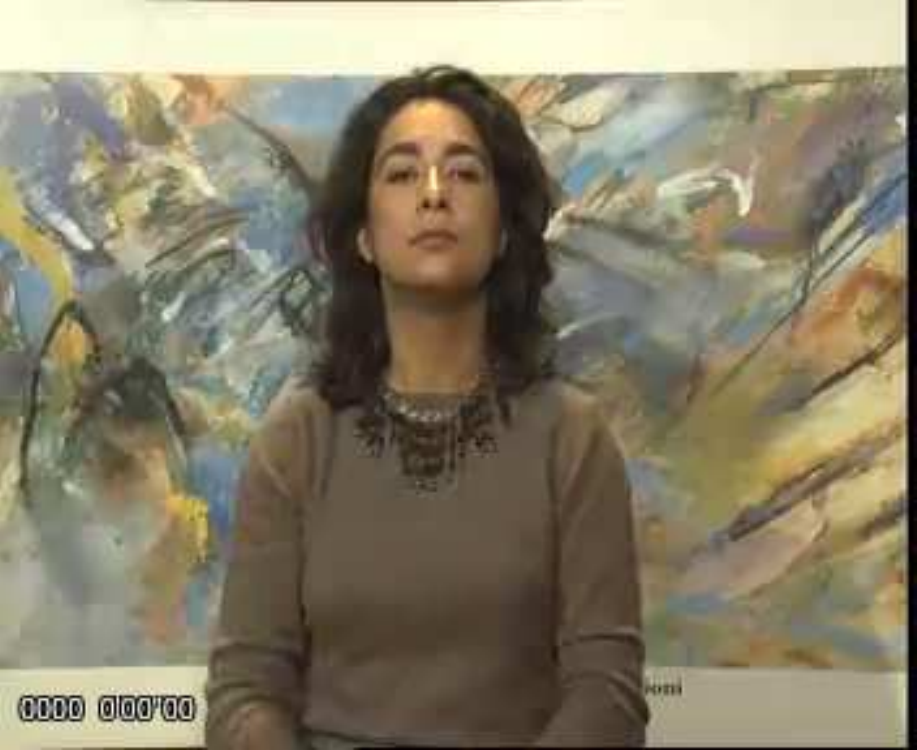}\label{fig:subfig8.3a}}\\
\subfigure[HEVC]
{\includegraphics[width=.32\linewidth]{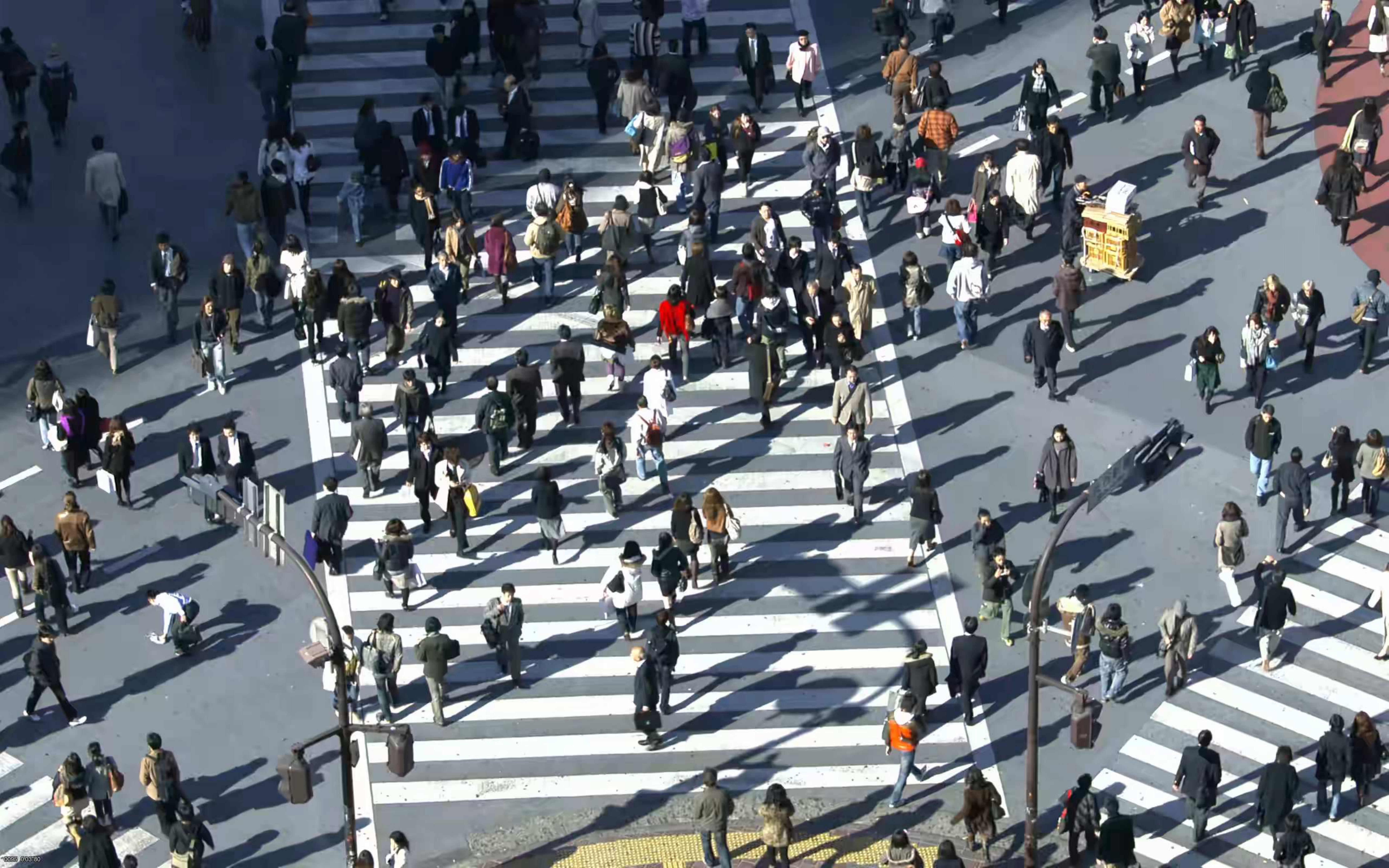}\label{fig:subfig8.1b}}
\subfigure[$\widetilde{\mathbf{C}}_N$]
{\includegraphics[width=.32\linewidth]{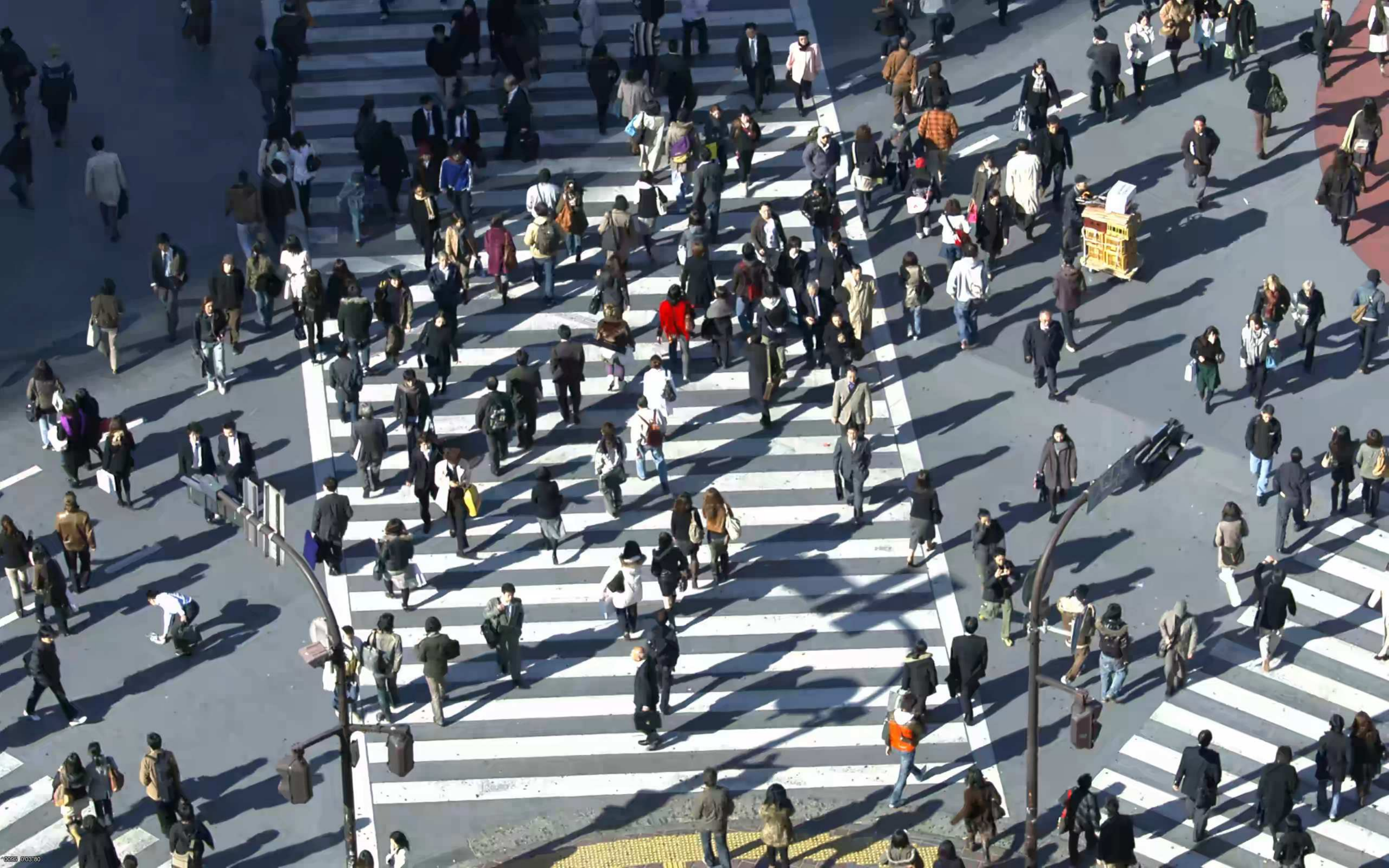}\label{fig:subfig8.2b}}
\subfigure[$\widehat{\mathbf{C}}_N$]
{\includegraphics[width=.32\linewidth]{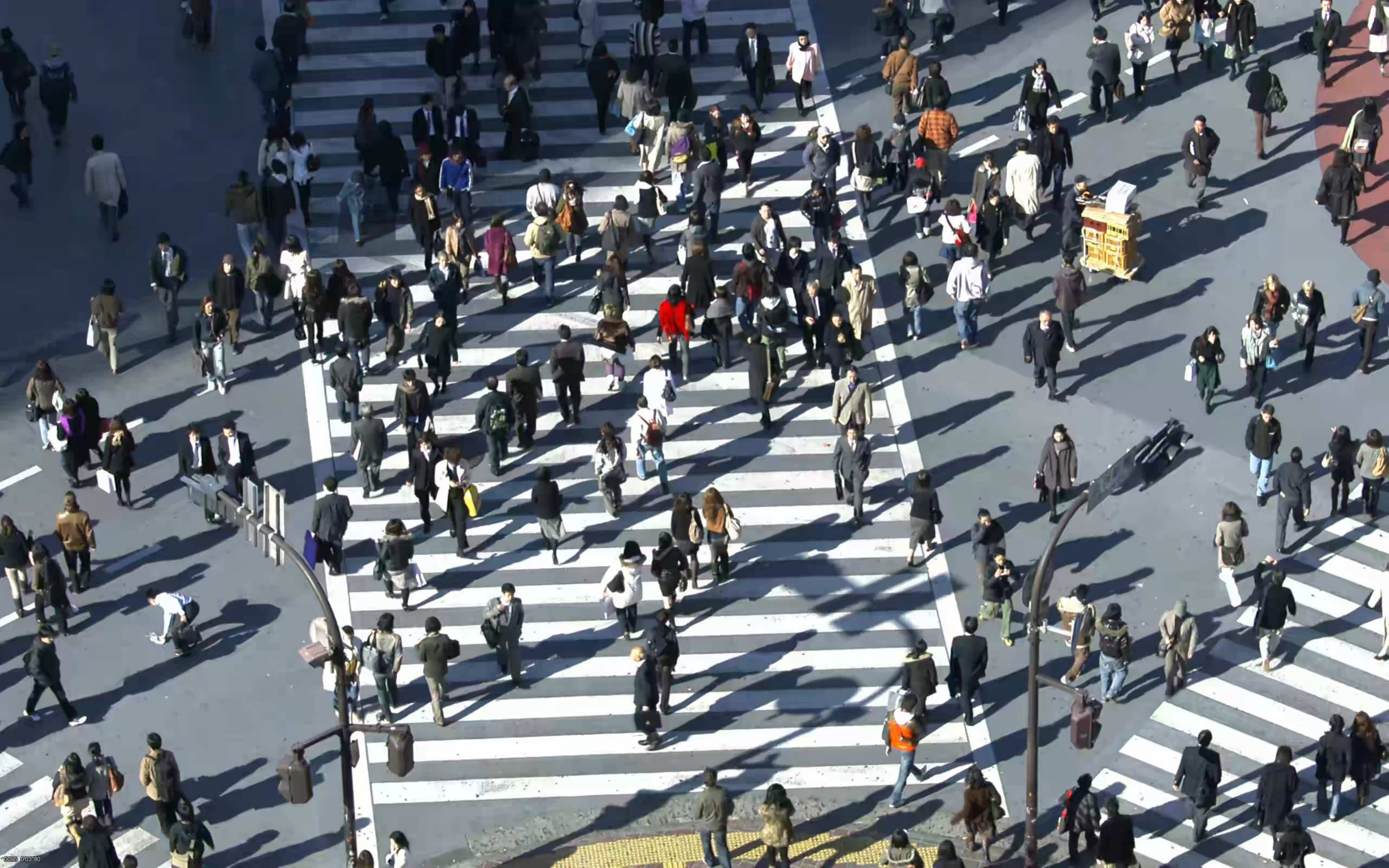}\label{fig:subfig8.3b}}
\caption[]
{
Qualitative comparison of frames from \texttt{Silent} and \texttt{PeopleOnStreet} videos compressed with
proposed Chen's DCT approximations
and default HEVC transforms and QP=32.}
\label{fig:figura8}
\end{figure}

As a qualitative example, Figure~\ref{fig:figura8}
displays particular frames of \texttt{Silent} and \texttt{PeopleOnStreet} video
sequences
after
compression according to the original HEVC
and
to the modified versions of HEVC based on the proposed transforms.
Visual inspection shows no sign of
image quality degradation.

\section{Conclusion}
\label{s:conclusion}

We introduced two new multiplierless DCT
approximations based on the Chen's factorization.
The suggested approximations were assessed
and
compared with other well-known approximations.
The proposed transformation
$\widehat{\mathbf{C}}_8$ presented
low total error energy
and
very close similarity to the exact DCT.
Furthermore,
$\widehat{\mathbf{C}}_8$
presents
very close coding gain
when compared to the optimal KLT.
The approximation $\widehat{\mathbf{C}}_8$
outperformed
the SDCT, BAS, and HT
as tools for
JPEG-like
still
image compression
at
a lower computational cost.
Adapting the JAM scalable method,
we also proposed low-complexity Chen's DCT approximations
$\widetilde{\mathbf{C}}_N$ and $\widehat{\mathbf{C}}_N$,
were $N\geq16$ is a power of two;
we also provided
fast algorithms for their implementations.
The introduced
8-, 16-, and 32-point
approximations
were embedded into an HEVC reference software and
assessed for video compression.
Finally,
the proposed low-complexity transforms
are suitable for image and video coding,
being a realistic alternative
for efficient
and low complexity
image/video coding.

\section*{Acknowledgments}

This research was partially supported by
CAPES, CNPq, and FAPERGS,
Brazil.

{\small
\singlespacing
\bibliographystyle{siam}
\bibliography{referencia,tlts-references}
}

\end{document}